\documentclass[journal,onecolumn,draftclsnofoot]{IEEEtran}

\usepackage{epsfig,subfigure,endnotes,footnote,amsthm,amsmath,amssymb,graphicx,epstopdf,wrapfig}
\usepackage{algorithm,algorithmicx, algpseudocode}
\usepackage{color,xcolor,url}
\usepackage{balance}
\usepackage{paralist}
\usepackage[total={7.23in, 9.23in}]{geometry}
\usepackage{amsfonts}
\usepackage [english]{babel}
\usepackage [autostyle, english = american]{csquotes}

\newcommand{\eg}{{\it e.g.}, }
\newcommand{\ie}{{\it i.e.}, }

\MakeOuterQuote{"}

\begin{document}

%%%%%%%%%%%%%%%%%%%%%%%
% Title portion. Note the short title for running heads 

\title{A Formal Approach for Efficient Navigation Management of Hybrid Electric Vehicles\\ on Long Trips}

\author{Mohammad Ashiqur Rahman, Md Hasan Shahriar, Ehab Al-Shaer, and Quanyan Zhu 
\thanks{M. A. Rahman (marahman@fiu.edu) and M. H. Shahriar (mshah068@fiu.edu) are with the Department of Electrical and Computer Engineering, Florida International University, USA. E. Al-Shaer (ealshaer@uncc.edu) is with the Department of Software and Information Systems, University of North Carolina at Charlotte, USA. Q. Zhu (quanyan.zhu@nyu.edu) is with the Department of Computer and Electrical Engineering, New York University, USA.}}

\maketitle

%%%%%%%%%%%%%%%%%%%%%%%%%%%%%%%%%%%%%%%%%%%%%%%%%%%
\begin{abstract}
%A Plug-in Hybrid Electric Vehicles (PHEV) are gaining popularity due to their economic efficiency as well as their contribution to environmental preservation. PHEVs 
%allow the driver to use electric power exclusively for 30-50 miles of driving, and then switch to gasoline for longer trips. 
Plug-in Hybrid Electric Vehicles (PHEVs) are gaining popularity due to their economic efficiency as well as their contribution to green management. PHEVs allow the driver to use electric power exclusively for driving and then switch to gasoline as needed.  
The more gasoline a vehicle uses, the higher cost is required for the trip. However, a PHEV cannot last for a long period on stored electricity without being recharged. Thus, it needs frequent recharging compared to traditional gasoline-powered vehicles. Moreover, the battery recharging time is usually long, which leads to longer delays on a trip. Therefore, 
%for the deployment of PHEV technology,
it is necessary to provide a flexible navigation management scheme along with an efficient recharging schedule, which allows the driver to choose an optimal route based on the fuel-cost and time-to-destination constraints. In this paper, we present a formal model to solve this PHEV navigation management problem. 
%We implement this model using Satisfiability Modulo Theories (SMT). 
The model is solved to provide a driver with a comprehensive routing plan including the potential recharging and refueling points that satisfy the given requirements, particularly the maximum fuel cost and the maximum trip time. In addition, we propose a price-based navigation control technique to achieve better load balance for the traffic system. Evaluation results show that the proposed formal models can be solved efficiently even with large road networks. %with large numbers of gas and charging stations.
%\remark{Change the CCS concepts with suitable ones.}
\end{abstract}


%%%%%%%%%%%%%%%%%%%%%%%%%%%%%%%%%%%%%%%%%%%%%%%%%%%
\section{Introduction}
\label{Sec:Introduction}

Plug-in Hybrid Electric Vehicles (PHEV) have the potential of addressing contemporary and future environmental and economic challenges. In addition, they provide new possibilities for smart grid management and the integration of renewable energy sources into electricity networks~\cite{Hybridcars}. 
%A PHEV is typically equipped with a rechargeable battery and can use the electric energy stored in the battery as energy, as an alternative to traditional gasoline. This battery can be fully recharged by connecting it to the power grid with an extension cord. 
A PHEV is typically equipped with a battery that can be fully recharged by connecting it to the power grid with an extension cord. The stored electric energy is used as alternative to traditional gasoline, providing economic and environmental benefits. The cost for electricity to power a PHEV has been estimated to be less than one fourth of the cost of gasoline~\cite{Calcars}. 

According to the typical battery capacity of an Electric Vehicle (EV), a driver is allowed to use electric energy exclusively for 30 to 50 miles of driving before switching to gasoline for longer trips~\cite{Hybridcars}.
%An electric vehicle (EV) cannot travel long distance using the stored electricity without recharging its battery.
Thus, an EV needs frequent recharging in longer trips. Moreover, recharging takes more time than refueling.
%An electric vehicle (EV) needs frequent recharging for long distances, which leads to more time than usual refueling, resulting longer delays.
As a result, EVs suffer with excessive delays in long trips. 
However, unlike EVs, a PHEV can use gasoline when stored energy is fully consumed instead of frequently recharging the battery. %always recharging. 
Therefore, it is necessary to select the optimal route to the destination that provides an efficient recharging and refueling schedule plan for a PHEV in order to satisfy different constraints. Vehicle drivers usually have constraints on driving time as well as on the fuel cost for a trip. There can be additional requirements like needing to pass through some specific locations (\ie intermediate points of interest or via points).

%\comment{TODO}
In this work, we model the PHEV navigation management as a constraint-satisfaction problem. 
%We also present  a \emph{graph theoretical} proof show that this navigation management problem is \emph{NP-complete}. 
Satisfiability Modulo Theories (SMT)~\cite{BM09} to implement this model and obtain a satisfiable solution that includes the routing plan and the recharging and refueling points, satisfying the constraints on fuel cost, driving time, and intermediate points of interest. 
This work assumes that a service provider can execute this formal model to compute a navigation plan for a PHEV based on its driver's requirements and the information about the road network, the gas and charging stations, the gas and charging prices, etc. 
A PHEV can obtain a navigation plan from the service provider at the beginning of its trip. That plan can be updated on-demand according to the latest information on traffic and charging stations. 
%based on the status update on the charging stations and traffic. 
We also propose a navigation control technique that adjusts the charging prices of the stations at each time slot and (indirectly) distribute the PHEVs, being motivated by the lower prices, on the roads and in the charging stations. 
%
%Our main contributions in this work are as follows:
%\begin{enumerate}
%\item[(i)] We define the \emph{PHEV Navigation Management Problem} and show that the problem is \emph{NP-complete} using a \emph{graph theoretical} approach.
%\item[(ii)] We formally model the navigation management problem using SMT. We demonstrate the model using an example. We evaluate the efficiency of our proposed model by executing simulation experiments. 
%%Our model takes less than a minute for a road network consisting of 500 points and a similar number of charging and gas stations.
%\item[(iii)] We present a price-based navigation control technique to achieve better load balancing
%for the system.
%%\item In addition, this work contributes to the green management, as it inherently minimizes the usage of gasoline by efficiently managing the use of the recharging technology.
%\end{enumerate}   

Unlike our previous work~\cite{RahmanICCPS13}, where we addressed the PHEV navigation management problem on highways with only the battery recharging, in this paper we extend the model to provide a comprehensive navigation management. This extended design models a generic road system with numerous distinguishing characteristics: location point as the unit of the road system, refueling and recharging requirements, flexible amounts of recharging and refueling, time-varying queue lengths for waiting vehicles in charging stations, and time-varying average traffic speeds.
We also devise a mechanism of executing several instances of the proposed model parallelly to solve a problem. This parallelism significantly increases the scalability of the solution.
The evaluation results shows that it takes few seconds to solve the proposed model for a road system consisting of 1,000 location points and a proportional number of charging and gas stations. Our work contributes to environmental preservation, as it inherently minimizes gasoline use by efficiently managing the use of recharging and refueling installations.

This paper is organized as follows: We discuss the state of the art of PHEVs and the PHEV navigation management problem in Section~\ref{Sec:Background}. %, and its complexity. 
We present the formal model of the navigation problem along with an illustrative example in Section~\ref{Sec:Model}. In the following section, we propose the navigation control technique and the corresponding model. Section~\ref{Sec:Evaluation} presents evaluation results. In the following section, we briefly discuss several aspects of the proposed solution. In Section~\ref{Sec:Related},  we briefly discuss the related work. Section~\ref{Sec:Conclusion} concludes the paper. %and proposes future work. 

%\vspace{-3pt}
%%%%%%%%%%%%%%%%%%%%%%%%%%%%%%%%%%%%%%%%%%%%%%%%%%%
\section{Background and Motivation}%%\vspace{-3pt}
\label{Sec:Background}

This section presents necessary background of the navigation management problem for PHEVs. %and discuss its complexity.
%Lastly, we discuss the motivation of using SMT for solving our model.

%\vspace{-6pt}
%%%%%%%%%%%%%%%%%
\subsection{PHEVs for Long Distance Travels}
\label{SSec:PHEV}

%%%%%%%%%%%%%%%%%%%%%%%%
%\begin{figure}[t]
%%\vspace{-9pt}
%\begin{center}
%\includegraphics[scale=0.7,keepaspectratio=true]{Figures/Fig_DailyMiles_Pie.pdf}
%%\vspace{-6pt}
%\caption{Percentages of vehicles with respect to daily miles driven. For example, it shows that 13\% of the vehicles travel 40-50 miles each day, while 20\% of the vehicles travel more than 50 miles per day.}
%\label{Fig_DailyMiles}
%\end{center} 
%%\vspace{-12pt}
%\end{figure}
%%%%%%%%%%%%%%%%%%%%%%%%

EVs are gaining slowly but steady popularity in the market. 
%It is expected that more than 10 million pure and hybrid EVs will be in the United States (U.S.) by 2020~\cite{CleanReport}. 
An estimated number of 361,307 of plug-in EVs , including hybrids, were sold in the United States (U.S.) in 2018~\cite{InsideEVs, CleanReport}.
Different models of PHEVs exist on the roads~\cite{Hybridcars, Calcars}. These cars usually can operate in the range of 50 miles after a full charge as a pure EV. Due to the recent development in PHEV technology, few well-advanced PHEVs are released to the market. 
%For example, the Hyundai BlueOn~\cite{Hyundai} can go more than 80 miles per charge and can be recharged to 80\% of the capacity in 25 minutes.
For example, the Hyundai IONIQ Electric~\cite{Hyundai} can go more than 120 miles per charge and can be fully recharged in around 30 minutes.
%; the Nissan Leaf electric car [5] runs up to 160 km per charge and takes 30 minutes to charge 80\% of the capacity of the battery pack. 
Though most of the cars run trips within short distances, around 20\% of the cars in the U.S. travel more than 50 miles% (see Figure~\ref{Fig_DailyMiles})
~\cite{Samaras08}. 
%Figure~\ref{Fig_DailyMiles} shows the percentages of vehicles with respect to daily miles driven. 
United States has a very large road network. According to the statistics recently published by the Federal Highway Administration, the U.S. interstate highway system alone has a total length of 47,714 miles~\cite{FHWA}. 
%Figure~\ref{Fig_Highways}, for example, shows a map of the major U.S. routes/roads (\ie interstate highways along with country routes) in North Carolina, particularly the roads between two of its cities: Gastonia and Winston-Salem~\cite{GoogleMap}.
%PHEVs are used for not only short-distance trips, but also long-distance journeys. 
%
%%%%%%%%%%%%%%%%%%%%%%%
\begin{wrapfigure}{r}{0.6\textwidth}
%\vspace{-15pt}
\begin{center}
\includegraphics[width=0.6\textwidth]{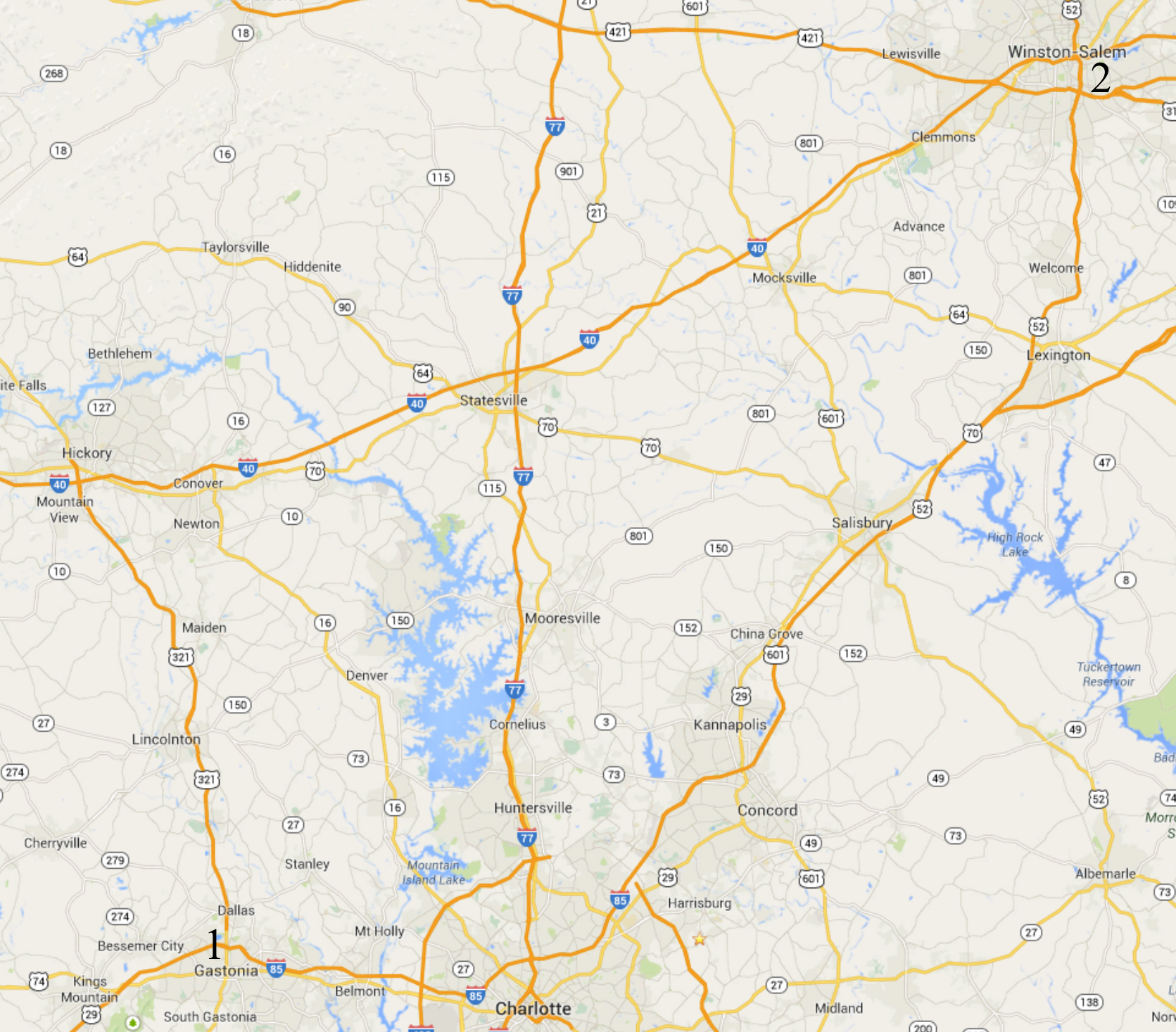}
\vspace{-12pt}
\caption{The map shows part of U.S. routes, particularly between two cities: Gastonia (Point 1) and Winston-Salem (Point 2) in North Carolina. The thick lines represent the roads.}
\label{Fig_Highways}
\end{center} 
\vspace{-9pt}
\end{wrapfigure}
%%%%%%%%%%%%%%%%%%%%%%%

PHEVs are used for long-distance and short-distance trips.
%In the cases of long trips, as a PHEV cannot run for long distances with the electric power only, the battery needs to be recharged to use electric power instead of gasoline for further operation. 
Since PHEVs cannot run for long distances with electric power only, the battery needs to be recharged to use electric power instead of using gasoline for further operation on long trips.
Again, PHEVs need long recharging times, which leads to long waiting times at charging stations. 
Figure~\ref{Fig_Highways} shows a map of the major U.S. routes/roads (\ie interstate highways along with country roads) between two cities, Gastonia and Winston-Salem, in North Carolina~\cite{GoogleMap}. 
As it can easily be seen in the figure that there are several alternative navigation paths from the source (point 1) to the destination (point 2). Therefore, there is always a need for choosing a suitable route.
For travel efficiency and comfortable driving, it is necessary to optimally plan a trip, \ie the routing path and the recharging and refueling schedule. %, so as to satisfy constraints on travel costs in terms of money and time.
There are few existing works related to the PHEV navigation management problem.  For example, a fuel-efficient navigation mechanism for conventional gasoline-powered vehicles is developed in~\cite{Raghu11}, while an optimal recharging scheduling for EVs is proposed in~\cite{Qin11}. However, neither of these model the uses of alternative fuels (electricity and gasoline) by PHEVs to satisfy the constraints for cost-effectiveness and time-efficiency.

%%%%%%%%%
\begin{figure}[t]
%\vspace{-15pt}
\begin{center}
\includegraphics[width=0.75\textwidth]{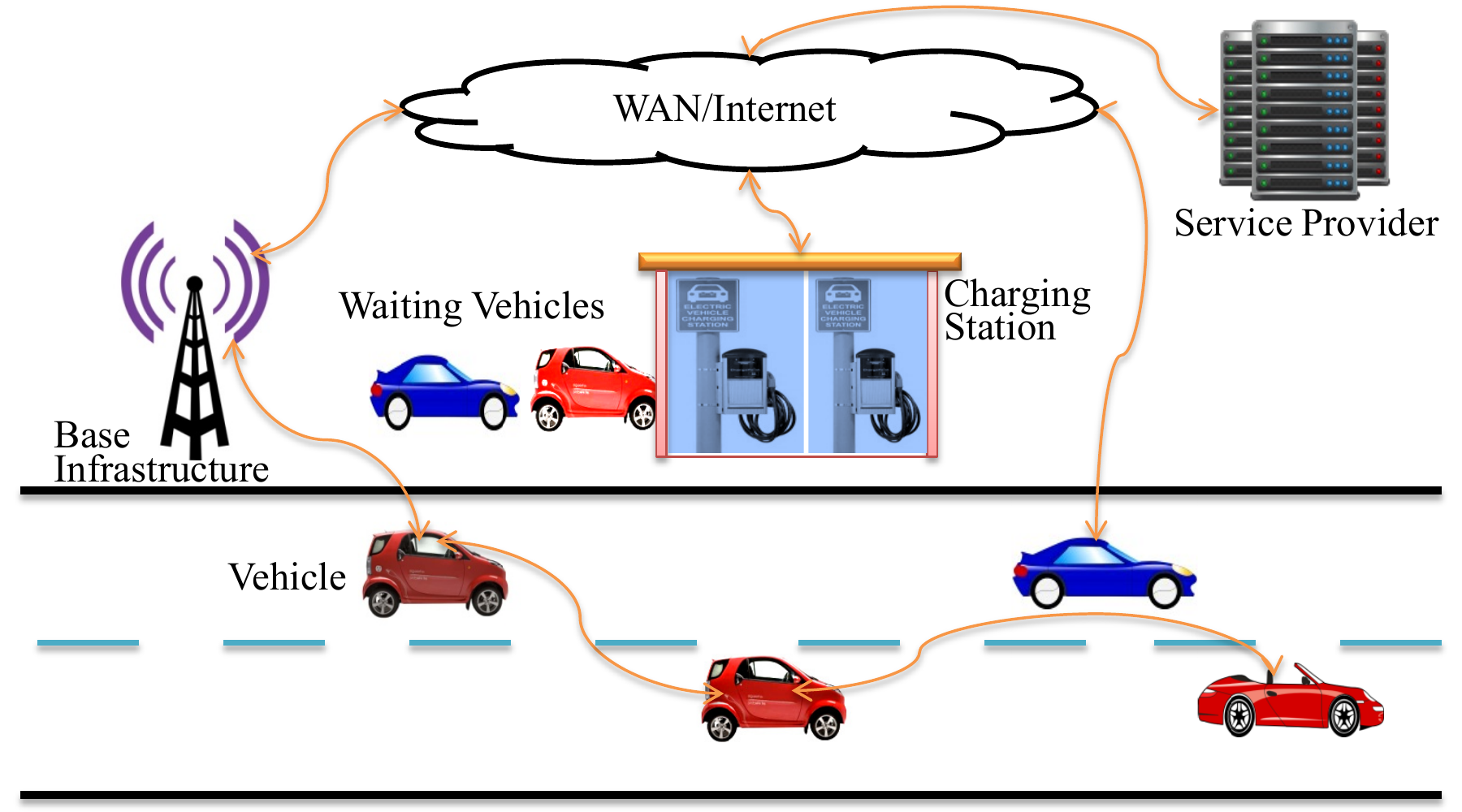}
%\vspace{-12pt}
\caption{The communication model for the navigation management system.}
%\caption{The illustration of the navigation control system}
\label{Fig_Comm}
\end{center} 
%\vspace{-12pt}
\end{figure}
%%%%%%%

In this paper, we assume a centralized system that provides a navigation plan to a PHEV for a long trip that satisfies the PHEV's (\ie its user's/driver's) requirements. The service provider will execute our navigation management model for synthesizing the navigation plan. However, a PHEV can also run the model by itself, if it has a processor for executing an SMT solver.
For navigation planning within the requirements, our model needs different static and dynamic information about the road network and corresponding traffic. 
%Hence, we assume a centralized/distributed information service existing for the highway system.
We assume that all the charging stations and the PHEVs are connected to the service provider through a wired or wireless, infrastructure-based or infrastructure-less communication model (\eg as shown in Figure~\ref{Fig_Comm}). The stations and PHEVs provide necessary local information to the service provider. 
%These servers can collect necessary data even from the road-side sensors, if they are deployed. 
The provider updates the dynamic part of the information system frequently with the collected data.  This dynamic information includes the current (real) status and future (predicted) status of the charging stations and the traffic on the road (especially, the number of waiting vehicles in the stations and the average traffic speeds on the roads).
The information specific to a future time is predicted from the past records and current status.
%, and this information is  for a period (\eg next 12 hours)
%A PHEV driver can provide the information to our model, which will generate a navigation plan for the vehicle.
%Figure~\ref{Fig_Comm} shows the general communication model that we assume in this work.
Since traffic can be very randomly distributed, spatially and temporally, the exact knowledge about the future %\ie the average vehicle speed on a road or the size of the waiting queue in a charging station 
may be different from the forecasted information. 
%Hence, the effectiveness of the navigation plan depends on the accuracy of the information. Since the nearer is a future time to the present, the more accurate is the prediction on that future. 
However, a PHEV can frequently update the navigation plan from the service provider based on the latest information and the same or modified requirements.

%%%%%%%%%%%%%%%%%%%%%%%%
%\begin{wrapfigure}{r}{0.6\textwidth}
%\vspace{-15pt}
%\begin{center}
%\includegraphics[width=0.6\textwidth]{Figures/Fig_Road_System.pdf}
%\vspace{-12pt}
%\caption{The communication model for the navigation management system.}
%%\caption{The illustration of the navigation control system}
%\label{Fig_Comm}
%\end{center} 
%\vspace{-12pt}
%\end{wrapfigure}
%%%%%%%%%%%%%%%%%%%%%%% 

%\vspace{-6pt}
%%%%%%%%%%%%%%%%%%%%%%%%%
\subsection{PHEV Navigation Management Problem}
\label{SSec:Problem}

The goal of the navigation management problem is to find an efficient navigation plan on a road system under a number of constraints. 
%Though the navigation management is the objective of the user (\ie driver) of the vehicle, for the ease of representation, we will consider vehicle as the subject instead of the user. 
Along with the basic objective of reaching the destination, on the way to the destination, a vehicle may require passing through one or more intermediate points of interest, \ie via-points. There are usually different time and cost constraints. There can be a time-to-reach-destination (or simply time) constraint, which means the vehicle should reach the destination on or before a specified time. There may also be time constraints to reach the via-points. 
The fuel-cost (or simply cost) incurred in a trip depends on the type and amount of the consumed fuel. A hybrid vehicle can use either electricity or gasoline as fuel. We assume that the vehicle does not use both kinds of fuel simultaneously, but rather sequentially. Hence, the cost is the summation of the price of electricity and that of gasoline consumed during the trip. The constraint on fuel-cost specifies that the cost cannot exceed a given value. 
%A PHEV usually requires to recharge its battery at the charging stations. 
We assume that a PHEV can recharge its battery at the charging stations only. 
Charging prices and waiting queues are usually different at different stations. Due to the time-varying price model of power-grids, a particular station may have different prices at different time slots. The objective of the PHEV navigation management problem is to find a driving routing plan from the source that potentially enables the vehicle to reach the destination, including the intermediate points of interest, within time and fuel-cost constraints. 
It was proven in~\cite{RahmanICCPS13} that the PHEV Navigation Management problem is NP-complete.

\subsection{SMT Logic and the Solver}
\label{SSec:SMT}

We use SMT to formalize our proposed navigation management models. SMT is a powerful logic tool that can solve constraint satisfaction problems arising in many diverse areas, such as software and hardware verification, test-case generation, scheduling, planning, etc.~\cite{MB09}. 
%An SMT instance is a formula in first-order logic, where some functions and predicate symbols have additional interpretations. 
SMT involves determining whether a formula is satisfiable or not. %~\cite{DP60}. 
For example, the SMT instance with the following two constraints is satisfiable with the assignments of $x=1$ and $y=0$:
\[ (x+y<2) \lor (x-2y>0)~~ \textrm{ and }~ x \leq 1\]
%This instance can be satisfied with the assignments: $x=0,y=0$. 

%SMT provides a much richer modeling language than is possible with SAT~\cite{Chaff}. 
An SMT instance is a first-order logic formula, where some functions and predicate symbols have additional interpretations. 
In SMT, complex logics are replaced by first order predicates/functions using a variety of underlying theories, including the theory of equality, linear arithmetic, difference logic, etc. 
%If a SMT instance cannot be satisfied, one needs to relax the constraints (\eg in our case, either increase the trip budget and/or the planned trip time) to find a satisfiable solution. 
An SMT solver searches for the solution(s) following an extended DPLL backtracking algorithm~\cite{Davis62, Nieuwenhuis06}.
Modern SMT solvers can check formulas with hundreds of thousands of variables and millions of clauses~\cite{MB09}. %Similar progress has also been observed for SMT solvers~\cite{MB09}.
In this work, we solve the proposed model using Z3, an efficient SMT solver, developed and managed by Microsoft Research~\cite{BM09, Z3}.
%Our evaluations show that the formalization can be solved for problems with thousands of (location) points and charging stations in the road system.

%%%%%%%%%%%%%%%%%%%%%%%
\begin{figure}[t]
%\vspace{-9pt}
\begin{center}
\includegraphics[width=0.7\textwidth]{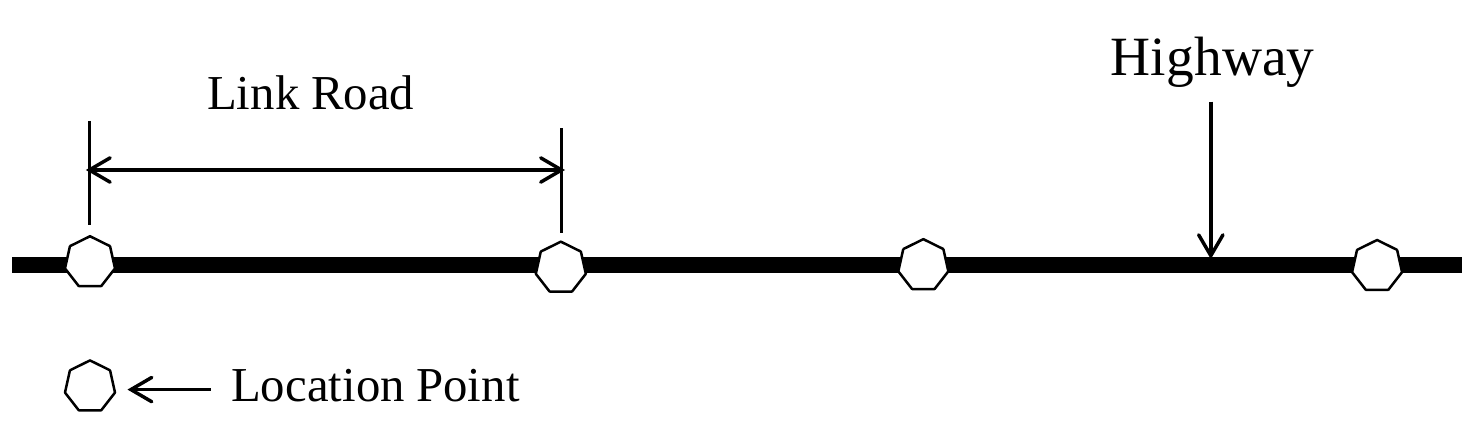}
\vspace{-6pt}
\caption{A route is modeled as a collection of small link roads, where a link road connects two location points.}
\label{Fig_Highway_Model}
\end{center} 
%\vspace{-12pt}
\end{figure}
%%%%%%%%%%%%%%%%%%%%%%%

%\vspace{-3pt}
%%%%%%%%%%%%%%%%%%%%%%%%%%%%%%%%%%%%%%%%%%%%%%%%%%%
\section{Navigation Management Model}%\vspace{-3pt}
\label{Sec:Model}

In this section, we formalize the PHEV navigation management problem as a satisfaction of a number of constraints. %We start the section by describing the highway system and its modeling.

%\vspace{-6pt}
%%%%%%%%%%%%%%
\subsection{Road Network}
\label{SSec:Road}

%\comment{Change the exit/entrance based highway model to a general road model.}
The roads used for long trips often are highway and country routes. These routes connected to each other directly (through ramps) or through other highways, country, or local roads and form the road network. We will use the term ``routes'' frequently to represent both highways and  country/local roads in the road network. 
A route usually consists of a large number of location points (simply locations or points), at which it connects with other routes/roads, or there are establishments of gas stations or charging stations, including the source, destination, and various places of interest.
%Figure~\ref{Fig_Highways} shows a map of the highways around the city of Charlotte (courtesy of \emph{Google Maps} \cite{GoogleMap}). 
Therefore, we can consider a route/road as a sequence of small roads, where each of these small roads links two points as shown in~Figure~\ref{Fig_Highway_Model}.
%Since an exit is co-located with an entrance point, we consider both as a single point and use the word \emph{exit point} (or simply \emph{point}) to denote both of them.
%A \emph{link road} may connect an exit of a highway to an entrance of another highway. 
We assume that all the roads are bidirectional. We do not consider the overhead of any traffic jam or signaling system on the routes, rather the effects of their existence are reflected in the average vehicle speeds on the roads. %\comment{Then, why not considering this average speed as a time-variant value.}

The gas and charging stations are deployed at different points. A PHEV can charge its battery at a point if there is a charging station, and then continue moving toward its destination. Similarly, if there is a gas station at a point, the vehicle can refuel. %and then continue moving to its destination. 
The road network (\ie the routes, locations, and the placements of the stations) is known. 
Different gas or charging stations often have different prices for refueling or recharging. 
%Due to the time-varying pricing model of the electricity, 
Different times of the day also have different prices for recharging. We assume that a day is divided into 24 hourly slots. The charging price remains the same within a slot while it may vary between the slots. Usually recharging a PHEV takes a substantially long time and, thus, there are often vehicles waiting for recharging in a station. The queue size of the waiting cars can be different at different times of the day. For example, the number of waiting vehicles at the end of the day is often larger than the number at midnight. It is worth mentioning that if a point has multiple charging stations, we consider a single charging station, averaging their charging prices, queue lengths, and other parameters. We do the same for the gas station. %Similar to the hourly based charging prices, we assume hourly based queue sizes for a charging station.

%\vspace{-6pt}
%%%%%%%%%%%%%%
\subsection{System Model}
\label{SSec:SysModel}

In this subsection, we present the modeling of routes/roads, stations, the vehicle (\ie a PHEV), and its user's requirements (\ie user requirements). Table~\ref{Tab_Notation} shows the notations used in the modeling. 
%\comment{We should not use "the" before the notations. E.g., "highway $h$", not "the highway $h$".}

\vspace{0.05in}
\noindent \emph{Road Network.} 
As already mentioned earlier, we define the road network as a collection of routes/roads consisting of a number of location points and the connecting roads. Thus, a route/road is modeled as a sequence of  points and the roads connecting them. A location point (also, simply, location or point) is denoted by a number, $l$.
%taking the associated road and the location. That is, a point is denoted by a pair of parameters $l$, where $h$ is the road and $l$ is the location on the road. 
$L_{\hat{l}, l}$ represents the road between location $\hat{l}$ to location~$l$. $D_{\hat{l}, l}$ is the distance (miles) between location $\hat{l}$ to location~$l$. 
%\comment{We need to make it time variant} 
The average speed at which a vehicle moves on the road is $S_{\hat{l}, l, t}$ (miles/minute) during time slot $t$. The average speed comes from the speed limit of the road and/or the traffic on the road.

\vspace{0.05in}
\noindent \emph{Charging Stations.} 
$S_l$ is a Boolean parameter denoting whether there is a charging-station at point~$l$. It has different properties. The price of charging (per kWh) at a time slot $t$ is denoted by $Ps_{l,t}$ (dollars). The expected queue length of the waiting vehicles at time slot $t$ is $Qs_{l,t}$.
%\comment{Has it been addressed}. 
If a particular vehicle is considered, it takes time $Ts_l$ minutes to recharge its battery for each kWh charge. In order to estimate the waiting time in queue, it is assumed that each vehicle in queue takes $\hat{T}s_l$ minutes on an average to recharge its battery. 
%We  assume that if a battery is recharged, it is recharged until it has reached its full capacity.
We do not consider the number of charging outlets, since this can easily be reflected in the average time required for each waiting vehicle. 
%A charging station often has more than one charging outlets. $Os_l$ shows the number of outlets at station $S_l$.
%\comment{Why not considering the number of charging points at the station?}

%%%%%%%%
\begin{table}[!t]
%\vspace{-9pt}
\caption{Modeling Parameters} \label{Tab_Notation}
%\vspace{-12pt}
\centering
\scriptsize
\begin{tabular}{|p{0.5in}|p{4in}|p{0.5in}|}
\hline
\textbf{Notation}	& \textbf{Definition} & \textbf{Type}\\
\hline
%$h$ & A route/road. \\
$l$ & A location point on the road network. & Integer \\
$L_{\hat{l}, l}$ & If there is a road between $l$ and $\hat{l}$. & Boolean \\
$D_{\hat{l}, l}$ & Length of road $L_{\hat{l}, l}$. & Real \\
$S_{\hat{l}, l, t}$ & Average speed of a vehicle on road $L_{\hat{l}, l}$ during time slot $t$. & Real \\
$S_l$ & If there is a charging station at point~$l$. & Boolean \\
$Ps_{l,t}$ & Price of charging (per kWh) at time slot $t$ at station $S_l$. & Real \\
$Qs_{l,t}$ & Queue length of waiting vehicles at time slot $t$ at station $S_l$. & Integer \\
$Ts_l$ & Time for recharging of each kWh charge at station $S_l$. & Real \\
$\hat{T}s_l$ & Average time for recharging at station $S_l$. & Real \\
$G_l$ & If there is a gas station at point~$l$. & Boolean \\
$Pg_l$ & Price of gasoline (per gallon) at station $S_l$. & Real \\
$\hat{T}g_l$ & Average time for taking the gasoline at station $G_l$. & Real \\
$Sv$ & Starting (or current) location of the vehicle. & Integer \\
$Dv$ & Destination point of the vehicle. & Integer \\
$Cv$ & Capacity of the vehicle's battery. & Integer \\
$\hat{C}v$ & Capacity of the vehicle's gasoline tank. & Real \\
$Ev$ & (Initially) stored electric charge of the vehicle. & Real \\
$Pv$ & Price of each kWh stored electric charge of the vehicle. & Real \\
$Gv$ & (Initially) stored gasoline of the vehicle. & Real \\
$\hat{P}v$ & Price of each stored gallon gasoline of the vehicle. & Real \\
%$Re$ & Electricity consumption of the vehicle for each mile. & Real \\
$Re$ & The number of miles that a vehicle can get per kWh electric charge. & Real \\
$Rg$ & Gas consumption of the vehicle for each mile. & Real \\
$Pg$ & Price of gasoline per gallon. & Real \\
$Cp$ & Cost constraint of the vehicle. & Real \\
$Ct$ & Time constraint of the vehicle to reach the destination. & Real \\
$Ct_l$ & Time constraint of the vehicle to reach via point $l \in I$, $I$ is the set of via points. & Real \\
$X_l$ & Whether the vehicle reaches point~$l$. & Boolean \\
$Sc_l$ & Whether station $S_l$ is selected for charging. & Boolean \\
$Ce_l$ & Stored electricity of the vehicle when it reaches point~$l$. & Real \\
$\hat{C}e_l$ & (Effective) stored electricity of the vehicle when it leaves point~$l$. & Real \\
$Cg_l$ & Stored gasoline of the vehicle when it reaches point~$l$. & Real \\
$\hat{C}g_l$ & (Effective) stored gasoline of the vehicle when it leaves point~$l$. & Real \\
$Pe_l$ & Average price of each kWh of electric charge stored in the vehicle at point~$l$. & Real \\
$Pg_l$ & Average price of each gallon gasoline stored in the vehicle at point~$l$. & Real \\
$P_l$ & Cost spent by the vehicle to reach point~$l$. & Real \\
$T_l$ & Time spent by the vehicle to reach point~$l$. & Real \\
$St_l$ & Time slot at which time $T_l$ falls. & Integer \\
\hline
\end{tabular}
\normalsize
%\vspace{-12pt}
\end{table}
%%%%%%

\vspace{0.05in}
\noindent \emph{Gas Stations.} 
Similar to a charging station, we also define several parameters for a gas station. 
%Due to the high availability of gas stations, 
$G_l$ denotes whether point~$l$ has a gas station ($G_l$). We define $Pg_l$ as the price of each gallon of gasoline at this station. Since there are often more than one gas stations at a point and they often have different prices although very close to each other, we can consider all of these gas stations as a single station and the price as the average of their prices. We define $\hat{T}g_l$ as the average time of taking the gasoline. Since the gasoline pouring time per gallon is very small considering other overheads, we only consider an average time. Moreover, we do not consider any waiting queue for a gas station, because such a queue is rare to occur at a gas station.

\vspace{0.05in}
\noindent \emph{The Vehicle.} 
A plug-in hybrid vehicle $V$ has different properties: the current location $Sv$ (an entrance point),  the destination $Dv$ (a point), the stored electric charge $Ev$ (kWh) and its price $Pv$ (\$/kWh), the battery capacity $Cv$ (kWh), the electric energy consumption rate $Re$ (miles/kWh), the stored gasoline $Gv$ (gallon) and its price $\hat{P}v$ (\$/gallon), the gasoline capacity $\hat{C}v$ (gallon), and the gasoline consumption rate $Rg$ (miles/gallon). We assume that any vehicle drives on a road at the speed limit of the road (\ie $S_{\hat{l}, l, t}$). %The price of gasoline $Pg$ is assumed as constant in the system. 

\vspace{0.05in}
\noindent \emph{User Requirements.} We consider that the vehicle is required to satisfy two major constraints. The first constraint is on the cost $Cp$ (dollars); that is, the price of the fuel consumed by the vehicle should be within a cost bound. The second constraint is on the traveling time $Ct_{Dv}$ (minutes); that is, the vehicle has to reach the destination $Dv$ within a time limit. There can be more constraints. The vehicle might need to go via some intermediate points denoted as a set $I$. Moreover, a vehicle might be required to reach a via-point within a time constraint ($Ct_l$ for ${l \in I }$). There can be other arbitrary user requirements, \eg minimum electricity stored when the destination is reached.

%\vspace{-6pt}
%%%%%%%%%%%%%%%%%
\subsection{Vehicle Routing Model}
\label{SSec:Model_Routing}

%\comment{TODO: Equations need to be changed based on the changed parameters.}
The routing of a vehicle from a source to a destination is a sequence of moves from one point to another from the source to the destination. The cost and time incurred due to this sequence of moves must be within the cost and time constraints. We assume that the vehicle's primary energy source is the battery, \ie the electric charge, as it costs significantly less than the gasoline. Therefore, if the battery is out of charge, then gasoline is used as the energy source. 
%We model the vehicle movement, the charging, and the constraints in this section.

%%%%%%%%
%\subsubsection{Parameters for Modeling Vehicle Routing}
\subsubsection{Modeling Parameters}

The following parameters are used to model a vehicle's routing to a destination: 
%They are described below:
 
%\begin{itemize}
\begin{compactitem}
\item \emph{Visited point.} $X_l$ denotes that the vehicle reaches point~$l$, \ie it moves to point~$l$ on route $h$. It is a boolean value. If the vehicle reaches the destination, the goal is achieved (\ie no more movement is required). 
$X_{\hat{l}, l}$ denotes that the vehicle moves from $\hat{l}$ to $l$.
%location $\hat{l}$ on $\hat{h}$ to location~$l$ on highway $h$. %We assume that no point is revisited. 

\item \emph{Charging station selected for recharging.} $Sc_l$ denotes whether the vehicle recharges at the charging station located at point~$l$. %It can be true if there is a charging station at $l$, \ie $S_l$.

\item \emph{Gas station selected for refueling gasoline.} $Gc_l$ denotes if the vehicle refuels at the gas station located at point~$l$. %It can be true if there is a charging station at $l$, \ie $S_l$.

\item \emph{Stored electricity.} $Ce_l$ represents the stored electric charge of the vehicle at point~$l$. 

\item \emph{Effective stored electricity.} $\hat{C}e_l$ is the effective  stored electricity at point~$l$. Obviously, $\hat{C}e_l \ge Ce_l$. If the vehicle is recharged at the charging station at point~$l$, it is greater than $Ce_l$.

\item \emph{Stored gasoline.} $Cg_l$ is the stored gasoline of the vehicle at point~$l$. 

\item \emph{Effective stored gasoline.} $\hat{C}g_l$ is the effective stored gasoline at point~$l$. Obviously, $\hat{C}g_l \ge Cg_l$. If the vehicle is refueled from the gas station at $l$, it is greater than $Cg_l$.

\item \emph{Average electricity price.} $Pe_l$ represents the average price of each kWh of electric charge stored in the vehicle at point~$l$. 

\item \emph{Average gasoline price.} $Pg_l$ represents the average price of each gallon gasoline stored in the vehicle at point~$l$. 

\item \emph{Trip-cost.} $P_l$ is the cost that the vehicle has spent to reach point~$l$, starting from the source. 

\item \emph{Time-to-reach.} $T_l$ is the time at which the vehicle has reached point~$l$. 
%starting from the source. 

\item \emph{Slot of a time.} $St_l$ represents the time slot in which the time $T_l$ falls. 
%\end{itemize}
\end{compactitem}
%\vspace{0.1in}

%%%%%%%%
%\subsubsection{Modeling Vehicle Movement}
\subsubsection{Modeling Vehicle Movement}

A vehicle can move to point~$l$ from point $\hat{l}$, if the vehicle is in $\hat{l}$ (\ie it is already reached) and there is a road from $\hat{l}$ to $l$. 
The vehicle can move to $l$ from any of the neighboring points (\ie the points connected to $l$). This requirement is encoded as follows:
%The point~$l$ can be reachable from any of the points which are connected to it (\ie neighboring points).
\begin{equation}
X_{\hat{l}, l} \rightarrow~ X_{\hat{l}} \wedge L_{\hat{l}, l}
\end{equation}
\begin{equation}
X_l \rightarrow~ \bigvee_{\hat{l}} X_{\hat{l}, l}
\end{equation}
%
%There is no need to move from the destination or to the source. 
%%In order to reduce the search space, we consider these issues as constraints. 
%Hence, we have following constraints:
%\begin{equation}
%\forall_l~ (\hat{l} = Dv) \rightarrow~  \neg X_{\hat{l}, l}
%\end{equation}
%\begin{equation}
%\forall_{\hat{l}}~ (l = Sv) \rightarrow~  \neg X_{\hat{l}, l}
%\end{equation}

In this model, we assume that no point is revisited. Due to this assumption, it is obvious that, if there are some intermediate destinations (\ie via-points), they have to be on the way toward the final destination. If there is a requirement to backtrack a path to reach the final (or an intermediate) destination, which could be mandatory due to the road system, then the model will fail to find a routing plan. The following constraints ensure that no point is revisited: 
%\begin{equation}
% X_{\hat{l}, l} \rightarrow \bigwedge_{{\hat{h'},\hat{l'}} \neq {\hat{h},\hat{l}}} \neg{X^{\hat{h'},\hat{l'}}_{h,l}}
%\end{equation}
\begin{equation}
X_{\hat{l}, l} \rightarrow \bigwedge_{\hat{l'} \neq \hat{l}} \neg{X_{\hat{l'}, l}}
\end{equation}
%
%\begin{equation}
%X_{\hat{l}, l} \rightarrow \bigwedge_{{h',l'} \neq {h,l}} \neg{X^{\hat{h},\hat{l}}_{h',l'}}
%\end{equation}
\begin{equation}
X_{\hat{l}, l} \rightarrow \bigwedge_{l' \neq l} \neg{X_{\hat{l}, l'}}
\end{equation}

%%%%%%%%
\subsubsection{Modeling Battery Recharging}

The battery of the vehicle can be recharged at a location if the vehicle have reached the point plus there is a charging station, as shown in Equation~(\ref{Eq_IfCharge}). If the battery is recharged, the stored charge of the battery increases. %but the increases is limited to its capacity. 
If it is not recharged, then the effective charge remains the same. 
We assume that when a battery is recharged, it is always recharged in an integer amount (kWh) but no more than its capacity.
%its whole capacity is utilized for recharging, \ie the battery is always recharged fully up to its capacity.
\begin{equation}
\label{Eq_IfCharge}
Sc_l \rightarrow X_l \wedge S_l
\end{equation}
\begin{equation}
Sc_l \rightarrow  (\hat{C}e_l >  Ce_l) \wedge (\hat{C}e_l \le Cv)
\end{equation}
\begin{equation}
\neg Sc_l \rightarrow (\hat{C}e_l = Ce_l)
\end{equation}
The formalization of the stored electricity of the vehicle when it reaches a point is shown in Equation~(\ref{Eq_Charge}). The stored electricity at point~$l$ depends on the stored electricity at the last visited point (\ie $\hat{l}$) and the charge required to cross the distance between these points. %(\ie the length of the connecting road). 
\begin{equation}
\label{Eq_Charge}
%\begin{split}
 X_{\hat{l}, l} \rightarrow~ (B \rightarrow ((Ce_l = \hat{C}e_{\hat{l}} - C) \wedge (\hat{D}_{\hat{l}, l} = 0)))~ \wedge
(\neg{B} \rightarrow ((Ce_l = 0) \wedge (\hat{D}_{\hat{l}, l} = D_{\hat{l}, l} - D)))
%\end{split}
\end{equation}
In Equation~(\ref{Eq_Charge}), $C$ and $D$ denote $(D_{\hat{l}, l} / Re)$ and $(\hat{C}e_{\hat{l}} \times Re)$, respectively. $B$ stands for the Boolean result of $(\hat{C}e_{\hat{l}} \ge C)$, and $\hat{D}_{\hat{l}, l}$ represents the distance traveled by electric power from $\hat{l}$ to $l$.
%
%\comment{Update the average price of each kWh of stored electric charge.}
If the battery is recharged at point~$l$, then the average cost of each kWh of electric charge stored in the battery needs to be updated as follows:
\begin{equation}
%\begin{split}
 X_{\hat{l}, l} \rightarrow~
 (Sc_l \rightarrow (Pe_l = (Pe_{\hat{l}} \times Ce_l + 
Ps_{l,\hat{t}} \times (\hat{C}e_l - Ce_l)) / \hat{C}e_l)~ \wedge
(\neg Sc_l \rightarrow (Pe_l = Pe_{\hat{l}}))
%\end{split}
\end{equation}
In this equation, $\hat{t}$ stands for $St_l$.

%%%%%%%%
\subsubsection{Modeling Gasoline Refueling}

The vehicle can refuel at the gas station at a location provided that the vehicle reaches that point, as shown in~(\ref{Eq_IfFuel}). If the tank is refueled, the stored gasoline of the vehicle increases. If it is not refueled, then the effective stored gasoline remains the same. 
We assume that when the gas tank is refueled, it is always refueled in an integer amount up to its capacity.
%its whole capacity is utilized for recharging, \ie the battery is always recharged fully up to its capacity.
\begin{equation}
\label{Eq_IfFuel}
Gc_l \rightarrow~ X_l \wedge G_l
\end{equation}
\begin{equation}
Gc_l \rightarrow~  (\hat{C}g_l >  Cg_l) \wedge (\hat{C}g_l \le \hat{C}v)
\end{equation}
\begin{equation}
\neg Gc_l \rightarrow~ (\hat{C}g_l = Cg_l)
\end{equation}
%
%\comment{Need to work on the following equations carefully}
The formalization of the stored electricity of the vehicle when it reaches a point is shown in Equation~(\ref{Eq_Fuel}). The stored gasoline at point~$l$ depends on the stored gasoline at the previous visited point ($\hat{l}$) and the gasoline consumed to travel the distance between these points. Remember that the gasoline is used as the energy source only when the battery is out of charge.
\begin{equation}
\label{Eq_Fuel}
%\begin{split}
 X_{\hat{l}, l} \rightarrow~ ((\hat{C}g_{\hat{l}} \ge \hat{C}) \wedge (Cg_l = \hat{C}g_{\hat{l}} - \hat{C}))
%\end{split}
\end{equation}
In Equation~(\ref{Eq_Fuel}), $\hat{C}$ represents $((D_{\hat{l}, l} - \hat{D}_{\hat{l}, l}) / Rg)$.
%, and $\hat{B}$ denotes the Boolean result of $(\hat{C}_{\hat{l}} \ge C)$.
%
%\comment{Update the average price of each gallon of stored gasoline.}
If the vehicle takes gasoline at point~$l$, then the average cost of each gallon of electric charge stored in the battery needs to be updated as follows:
\begin{equation}
%\begin{split}
 X_{\hat{l}, l} \rightarrow~
(Gc_l \rightarrow (Pg_l = (Pg_{\hat{l}} \times Cg_l + 
Pg_l \times (\hat{C}g_l - Cg_l)) / \hat{C}g_l)~ \wedge
(\neg Gc_l \rightarrow (Pg_l = Pg_{\hat{l}}))
%\end{split}
\end{equation}

%%%%%%%%
\subsubsection{Modeling Time and Cost Spent}
\label{SSSec:Model_Time_Cost}

The computation of the cost of traveling from one point to another depends on the energy type, \ie electric charge and gasoline, that has been used by the vehicle. In a trip, the vehicle can use any one of them or both based on the availability of the battery charge. The cost, $P_l$, to reach point~$l$ is modeled in Equation~(\ref{Eq_Cost}) according to our assumption that the vehicle's primary energy source is the battery. In the equation, %$P_{\hat{l}}$ is the cost to reach point $\hat{l}$,  
$PG_l$ represents $(((D_{\hat{l}, l} - \hat{D}_{\hat{l}, l}) / Rg) \times Pg_{\hat{l}})$ that denotes the price of the gasoline consumed during traveling from $\hat{l}$ to $l$, and $PE_l$ stands for $((\hat{C}_{\hat{l}} - C_l) \times Pe_{\hat{l}})$ which denotes the price of the electric charge used during traveling from $\hat{l}$ to $l$.
%\comment{Why not considering the consumed electricity to calculate the cost of electricity?}
\begin{equation} 
\label{Eq_Cost}
X_{\hat{l}, l} \rightarrow (P_l = (P_{\hat{l}} + PG_l + PE_l))
\end{equation}
The time required for traveling does not depend on the energy source (fuel) type
%, as we assumed that the speed of the vehicle does not depend on the source 
(refer to Section \ref{SSec:SysModel}). It depends on the average vehicle speed on the road. Equation~(\ref{Eq_Time}) shows the modeling of time computation.
\begin{equation}
\label{Eq_Time}
%\begin{split}
 X_{\hat{l}, l} \rightarrow~
(Sc_{\hat{l}} \rightarrow (T_l = (T_{\hat{l}} + T + TQ)))~ \wedge
(\neg{Sc_{\hat{l}}} \rightarrow ((T_l = (T_{\hat{l}} + T))))
%\end{split}
\end{equation}
In the above equation, $T$ stands for $(D_{\hat{l}, l} / S_{\hat{l}, l, \hat{t}})$ and $TQ$ represents $(Qs_{\hat{l},\hat{t}} \times \hat{T}s_{\hat{l}} + (\hat{C}_{\hat{l}} - C_{\hat{l}}) \times Ts_{\hat{l}})$.

%%%%%%%%
\subsubsection{Modeling User Requirements}

The main objective of the model is to find a route from the source to the destination. There are three more user requirements on the navigation plan: limited cost ($Cp$), time boundary ($Ct$), and the intermediate points of interest ($I$) with/without associated time boundaries ($Ct_l$ for ${l \in I }$). The following equations represent these constraints:
\begin{equation}
X_{Sv} \wedge X_{Dv} \bigwedge_{l \in I} X_l
\end{equation}
\begin{equation}\label{Eq_cost_constr}
P_{Dv} \le Cp
\end{equation}
\begin{equation}\label{Eq_time_constr}
T_{Dv} \le Ct
\end{equation}
\begin{equation}
\forall_{l \in I } T_l \le Ct_l
\end{equation}
It is easy to understand that, in the navigation plan, a point cannot be reached by the vehicle taking more cost or time than the overall cost or time boundaries (\ie $Cp$ or $Ct$), respectively. We incorporate these constraints as follows:
\begin{equation}\label{Eq_cost_constr2}
 X_l \rightarrow P_l \le Cp
\end{equation}
\begin{equation}\label{Eq_time_constr2}
X_l \rightarrow T_l \le Ct
\end{equation}
The user may have other constraints to satisfy. For example, when the vehicle reaches the destination, the stored electricity in the battery may need to be at least to some threshold value, which can be used for the next trip. This constraint is formalized as follows:
\begin{equation}
\hat{C}_{Dv} \ge Ce
\end{equation}
Here, $Ce$ is the threshold value.

%%%%%%%%
\subsubsection{Modeling Initial Constraints}

We need to consider the constraints that initialize the model. These constraints are about the stored electric capacity, the time and the cost at the source point. The starting time of travel (\ie time at source) is initialized with the current time. Equation~(\ref{Eq_Init}) models the initialization constraints.
\begin{equation}
\label{Eq_Init}
%\begin{split}
(Ce_{Sv} = Ev) ~\wedge (Cg_{Sv} = Gv) ~\wedge
(T_{Sv} = T_{current}) ~\wedge  (P_{Sv} = 0)
%\end{split}
\end{equation}
We need to initialize the prices for each kWh of stored electric charge and each gallon of stored gasoline at the source point ($Sv$). The initialization depends on whether the battery is recharged at this point. This is formalized as follows:
\begin{equation}
%\begin{split}
(Sc_{Sv} \rightarrow~ (Pe_{Sv} = (Pv \times Ce_l + 
Ps_{Sv,\hat{t}} \times (\hat{C}e_{Sv} - Ce_{Sv})) / \hat{C}e_{Sv}) ~\wedge
(\neg Sc_l \rightarrow (Pe_{Sv} = Pv))
%\end{split}
\end{equation}
Similarly, the price of each gallon of stored gasoline at point $Sv$ is initialized as follows:
\begin{equation}
%\begin{split}
(Gc_{Sv} \rightarrow~ (Pg_{Sv} = (\hat{P}v \times Cg_l + 
Pg_{Sv} \times (\hat{C}g_{Sv} - Cg_{Sv})) / \hat{C}g_{Sv})~ \wedge
(\neg Gc_{Sv} \rightarrow (Pg_{Sv} = \hat{P}v))
%\end{split}
\end{equation}

%%%%%%%%%%%%%%%
%\subsection{Implementation}
%\label{SSec:Satisfaction}
%
%We implement our model by encoding the system configuration and the constraints into SMT using \emph{Z3} SMT solver~\cite{Z3}. 
%We use the \emph{Z3 .Net API} for encoding the formalization of the model that we have already described in this section. 
%%which follows the same model that is already described in this section. 
%We use three types of terms: \emph{boolean}, \emph{integer}, and \emph{real}. 
%Though many parameters (\eg distance, speed, queue size, etc) in the model are usually integer values, we encode them as real terms, since we need to do some division operations in computing the time and the cost incurred for the vehicle routing.
%The system configurations and the constraints are given in a text file (\emph{input} file). The inputs are parsed into the model.
%%
%Executing the model (in Z3), we obtain the verification result as either satisfiable (\emph{sat}) or unsatisfiable (\emph{unsat}). If the result is \emph{unsat}, it means that the problem has no route from the source to the destination satisfying the constraints. In the case of \emph{sat}, we get the navigation plan from the assignments of the variables, $X_l$ and $Cs_l$. $X_l$ shows the route, while $Cs_l$ represents the station that is selected for recharging. The results from our model is also printed in a text file (\emph{output} file).

%\vspace{-6pt}
%%%%%%%%%%%%%
\subsection{A Case Study}
\label{SSec:Case}

%%%%%%%%%%%%%%%%%%%%%%%%
\begin{wrapfigure}{r}{0.6\textwidth}
\vspace{-36pt}
\begin{center}
\includegraphics[width=0.6\textwidth]{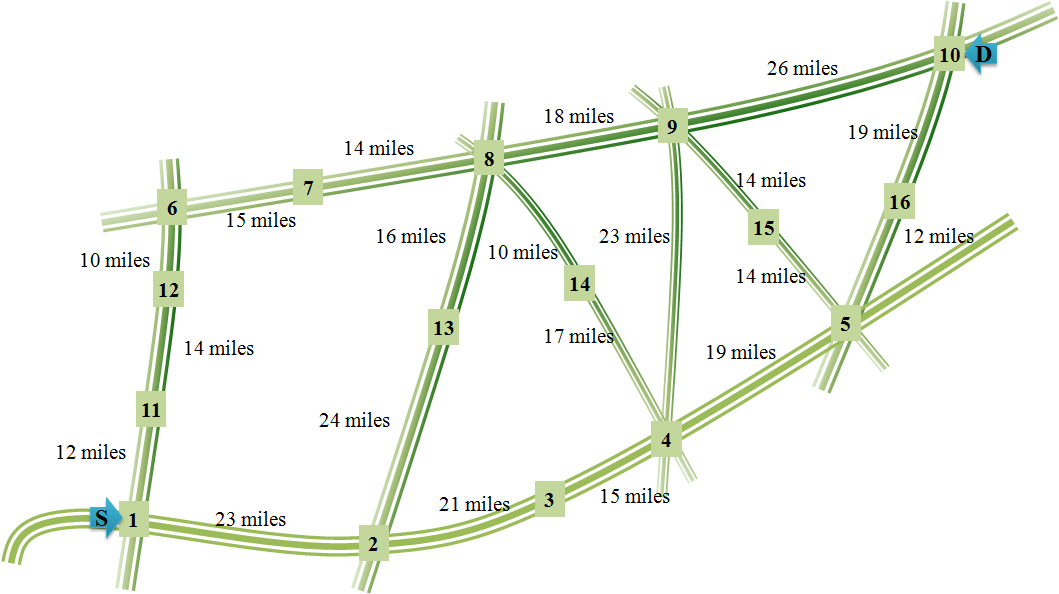}
\vspace{-18pt}
\caption{The topology of the road network considered in the example illustrated in Section~\ref{SSSec:Example}. This topology resembles the major routes from Gastonia to Winston-Salem. The locations are numbered from 1 to 16, while the source and destination of the trip are specified using two arrows.}
\vspace{-18pt}
\label{Fig_Example}
\end{center} 
\end{wrapfigure}
%%%%%%%%%%%%%%%%%%%%%%%%
%
Here, we first briefly discuss the implementation of the proposed formal model. Then, we present a synthetic case study demonstrating the execution of the model.

%%%%%%%
\subsubsection{Implementation}
\label{SSSec:Implementation}

We implement our model by encoding the system configuration and the constraints into SMT using the Z3 SMT solver~\cite{Z3}. 
%We use the \emph{Z3 .Net API} for encoding the formalization of the model that we have already described in this section. 
%which follows the same model that is already described in this section. 
The encoding requires to use Boolean, integer, and real terms. 
Although many parameters like distance, speed and queue size usually take integer values, we encode them as real terms, since some division operations that generate fractions are needed to perform.
The system configurations and the constraints are given in a text file. The inputs are parsed into the model.
Executing the model (in Z3), we obtain the verification result as either satisfiable (SAT) or unsatisfiable (UNSAT). If the result is UNSAT, it means that the problem has no route from the source to the destination satisfying the constraints. In the case of SAT, we get the navigation plan from the assignments of the variables, $X_l$, $\mathit{Sc}_l$, and $\mathit{Gc}_l$. $X_l$ shows the route, while $\mathit{Sc}_l$ and $\mathit{Gc}_l$ represent the stations that are selected for recharging and refueling, respectively. %The results from our model is also printed in a text file.

%%%%%%%%%%%%
\begin{table}[!t]
%\vspace{-9pt}
%\begin{singlespace}
\caption{Input to the Example} \label{Tab_Input}
%\vspace{-12pt}
\centering
\scriptsize
\begin{tabular}{|p{5in}|}
\hline

\# The numbers of locations and time slots \\
16 8

\vspace{0.05in}
\# Links (location pair, length, avg traffic speed) \\
20\\
1 2	23 1\\
2 3	21 1\\
3 4 15 1\\
4 5 19 1\\
6 7 15 1\\
%7 8 14 1\\
%8 9 18 3/4\\
%9 10 26 3/4\\
%1 11 12 3/4\\
%11 12 14 3/4\\
%12 6 10 3/4\\
%2 13 24 3/4\\
%13 8 16 3/4\\
%4 14 17 3/4\\
%14 8 10 1\\
%4 9 23 1\\
%5 15 14 1\\
%15 9 14 1\\
%5 16 12 1\\
%16 10 19 1
\ldots \ldots \ldots

\vspace{0.05in}
\# Charging stations (location, charging time/kWh, charging \\
\# time/waiting vehicle, queue size, price/kWh at each time slot) \\
6 \\
  1 2 20 4    1 3/2 3/2 5/4 5/4 5/4 3/4 3/4\\
12 2 15 1    1 3/2 3/2 3/2 5/4 5/4 3/4 3/4\\
13 2 18 2 3/4    1 3/2    1 5/4 5/4 3/4 3/4\\
  9 2 16 1 3/4    1 3/2 3/2    1 5/4 3/4 3/4\\
  4 2 18 2 5/4 3/2 3/2 3/2 5/4 5/4 3/4 3/4\\
16 2 16 2    1 3/2 3/2 5/4 5/4 5/4 3/4 3/4
       
\vspace{0.05in}
\# Gas stations (location, price/gallon, avg fueling time) \\
16\\
1 4 8\\
2 39/10 10\\
3 38/10 10\\
4 38/10 10\\
5 39/10 10\\
6 39/10 8\\
%7 38/10 8\\
%8 4 10\\
%9 39/10 10\\
%10 4 10\\
%11 39/10 10\\
%12 38/10 10\\
%13 38/10 10\\
%14 36/10 8\\
%15 36/10 10\\
%16 38/10 12
\ldots \ldots \ldots

\vspace{0.05in}
\# PHEV's properties (source, destination, stored energy and its price/kWh, battery capacity, \\
\# stored gas and its price/gallon, gas capacity, kWh/mile, gallon/mile, trip start time) \\
1 10 2 5/4 8 0 0 10 1/8 1/12 0

\vspace{0.05in}
\# Constraints (time, price) \\
130 21

\vspace{0.05in}
\# Via points \\
0 
\\ \hline
\end{tabular}
\normalsize
\vspace{-9pt}
%\end{singlespace}
\end{table}
%%%%%%%%%%%%

%%%%%%%
\subsubsection{An Example Case Study}
\label{SSSec:Example}

%\comment{The example needs to be a good one.}
We demonstrate our model through a synthetic case study. Figure~\ref{Fig_Example} shows the topology of the road network considered in this study. The topology resembles the major routes between Gastonia and Winston-Salem, two cities in North Carolina. The input file regarding this example is shown in Table~\ref{Tab_Input}. The objective of the PHEV is to find a routing plan from Gastonia (location 1) to Winston-Salem (location 10) that satisfies the given constraints. The vehicle starts its trip at time 0 and, according to the constraints, the vehicle needs to reach the destination by time 130 (130 minutes) and with no more than \$21 of energy cost. The vehicle has no via-point to pass by. 
The execution of the model corresponding to the example returns a SAT result. According to the solved result, the assignments of values to different variables of the model, we find that a possible route for the vehicle is \{1, 11, 12, 6, 7, 8, 9, 10\}. The battery needs to be recharged at the station at location 12, and the vehicle is refueled at location 1. The energy cost required through this route is \$20.83, while the time cost is \$129.5 minutes, both of which satisfy the constraints. It is worth mentioning that if the trip start time is changed to 60, then there is no routing path from the source to the destination under the above properties and constraints. In this case, if we can increase the time-to-reach constraint to time 220 (\ie 160 minutes of traveling time considering the start time) and increase the trip cost to \$22, then there is a solution, wherein the vehicle needs to use more electric charge compared to the previous case in order to keep the cost low. As a result, the vehicle needs to recharge for twice (at locations 12 and 9). The reason for this difference is that the charging price is higher during the trip time.

Let us go back to the first scenario, except with the following changes, as the vehicle is now more sensitive to cost compared to time. In this case, the trip cost cannot be more than \$18 while the vehicle must reach the destination by time 160 (160 minutes). We receive a satisfiable answer with the following results: the routing path is \{1, 11, 12, 6, 7, 8, 9, 10\}. The vehicle needs to recharge at locations 12 and 9, while refueled at location 1. The vehicle uses electric charge to travel the largest part of the route in this trip.

%\vspace{-6pt}
%%%%%%%%%%%%%%%%%%%%%%%%
\subsection{Optimal Navigation Plan Determination} 
\label{SSec:Optimal}

The verification result comprehensively represents a consistent PHEV navigation plan for the network, satisfying the user requirements. Usually, there are more than one model that satisfy the constraints. These models also take different time and cost, though all of them take time and cost less than or equal to the time constraint ($Ct$) and the cost constraint ($Cp$). 
Observing these models, one can choose the most cost- (or time-) efficient route among all alternative satisfiable models for the same set of constraints. %based on his or her criteria (\ie minimum time or minimum cost). 
We propose Algorithm~\ref{Algo_Optimal_Config} that 
%finds the optimal navigation plan based on the cost. The algorithm 
considers two values: $Cp_{max}$ and $Cp_{min}$, and finds the optimal navigation plan based on the cost. Typically, $Cp_{max}$ is the user's given constraint $Cp$, while $Cp_{min}$ is zero.
%represents the cost constraint $Cp$ as given by the user, while $Cp_{min} = 0$.
%The algorithm utilizes a binary search method, to find the optimal value. It is worth mentioning that Algorithm~\ref{Algo_Optimal_Config} may take a longer time, since the algorithm requires several invocations for the model synthesis.

\begin{algorithm}[t]
\caption{Optimal Configuration Determination}
\label{Algo_Optimal_Config}
\small
\begin{algorithmic}[1]
%\State $A$ is the property for optimization.
%\State $Th_A$ is the constraint (limit) value corresponding to $A$.
\State $Cp_{min}$ \& $Cp_{max}$ are the minimum and maximum values of $Cp$.
%\State $Th_{A,max}$ is the maximum possible $Th_A$.
\State $K_{max}$ is the maximum number of executions of the loop-body. 

%\Comment{Form the verification result we update $Th_{A,max}$. If the solver returns UNSAT, we update $Th_{A,min}$.}
\If{Solver returns SAT}
	\State Get Model, $\mathcal{M}$.
	\State Update $Cp_{max}$ according to the value of $P_{Dv}$ in $\mathcal{M'}$.
	\State $K = 0$.
	\Repeat
		\State $Cp = (Cp_{min} + Cp_{max})/2.$		
		\State Update the constraints~(Equations~(\ref{Eq_cost_constr}) and~(\ref{Eq_cost_constr2})) associated to $Cp$.		
		\If{Solver returns SAT}
			\State Get Model, $\mathcal{M'}$.
			\State Update $Cp_{max}$ according to the value of $P_{Dv}$ in $\mathcal{M'}$.
		\Else
			\State $Cp_{min} = Cp$.
		\EndIf
		\State $K = K + 1$.
	\Until {($Cp_{max} - Cp_{min} \approx 0$) \textbf{or} ($K = K_{max}$).}
	%\State Print $Th_{A,min}$.
\EndIf
\end{algorithmic}
%\vspace{-3pt}
\normalsize
\end{algorithm}

The algorithm utilizes a binary search method to find the optimal value. Algorithm~\ref{Algo_Optimal_Config} usually takes a longer time compared to the time for finding a satisfiable model only, since the algorithm requires several invocations of the model synthesis. The complexity of the algorithm is $O(T_{verify} \times {{log}_2}{D})$, where  $T_{verify}$ is the synthesis time and  $D$ is the difference between $Th_{A,max}$ and $Th_{A,min}$. Since $T_{verify}$ is very high in unsatisfiable cases as well as in tight constraint-based cases (see Section~\ref{Sec:Evaluation} for details), the time for finding the optimal would be very high for a large number of vehicles. However, if the navigation management service is provided online, where the formal model is executed by an SMT solver running in a centralized server, the navigation management service provider can utilize powerful machines to compute this optimization. The algorithm also lets the provider control the number of steps ($K_{max}$) to reduce the optimization time. In this case, a vehicle may receive a semi-optimal navigation plan.

\section{Price-based Navigation Control}%\vspace{-3pt}
\label{Sec:Control}

%The navigation (route) planning model provides a single stage solution (a route) that satisfy the time and cost constraints. Since we consider long trips, over the time the predicted traffic scenarios may change, causing traffic congestions on the overly chosen roads. The idea of navigation control technique is to distribute the vehicles on the roads adopting an incentive process (changing the electric charging prices) which can lead a lesser traffic congestion scenario. When the prices are changed, some of the existing vehicles’ current routing paths may not satisfy the respective drivers’ time and cost requirements any more for the rest of the trips. Hence, such a vehicle needs a new navigation plan for the remaining distances. These new routing paths, along with the paths for the incoming vehicles on the road system, ultimately help to control the traffic congestion. A driver may change the constraints to continue to follow the initial/current routing path.

The model we have discussed in Section~\ref{Sec:Model} finds a satisfiable navigation plan for a PHEV under a number of given constraints. The main constraints are time and cost.
The traveling cost depends not only on the electricity price but also on the availability of the charging stations, the queue lengths of waiting vehicles in the stations, the time constraint to reach the destination, and the vehicle speeds on different roads. The queue lengths at charging stations and the vehicle speeds on different road segments (which incorporate the traffic congestion) at different time slots are predicted from historical/past data and current/present status. Since we consider long trips, the initially predicted traffic scenario may change during the trip period. Moreover, due to non-speculated incidents, the number of vehicles choose a route or route segments can be larger than estimated. While heavy traffic can slow down the vehicles than the speed limits, the queue sizes of the charging stations on these routes have much higher possibility of being longer compared to that of other stations. While some routes may become busier than forecasted, many other routes can become freer than expected. Therefore, to deal with the uncertainty in predicted data, it is advantageous to control the (unexpected) traffic flow by distributing the unexpected loads among the roads, so that the roads have lesser traffic jams and the charging stations have smaller waiting queues. In this section, we present a simple motivation-based load control technique that motivates the cost-sensitive users to choose the roads with fewer loads.
The navigation management service provider can play the role of a controller in this technique.
However, the success of the model depends on the even distribution of the charging stations throughout the system. %, we assume that the charging stations are evenly distributed throughout the system.

%The model we have discussed in Section~\ref{Sec:Model} finds a satisfiable navigation plan for a PHEV under a number of given constraints. The main constraints are time and cost. The traveling cost depends not only on the electricity price but also on the availability of the charging stations, the queue lengths of waiting vehicles in the stations, and the time constraint to reach the destination. If many vehicles choose the same routes or route segments (\ie roads), the queue sizes of the charging stations on these routes have much higher possibility of being longer compared to that of other stations. Moreover, a large number of vehicles may create traffic jams on these roads.
%%, even though these roads are highways. We assume in our model that there would be no traffic jam on highways. 
%Hence, it is important to control the traffic flow for distributing the load among the roads, so that the roads have less or no traffic jams and the charging stations have smaller waiting queues. 
%%Though a controller cannot fully ensure these requirements, it can give an effort to accomplish them.
%In this section, we present a simple load control technique, which motivates the navigation management model to choose the roads with lower loads. 
%The navigation management service provider can play the role of a controller.
%However, the success of the model depends on the even distribution of the charging stations throughout the system. %, we assume that the charging stations are evenly distributed throughout the system.

%%%%%%%%%%%%%%%%%%%%%%%
\begin{figure}[t]
%\vspace{-9pt}
\begin{center}
\includegraphics[width=0.75\textwidth]{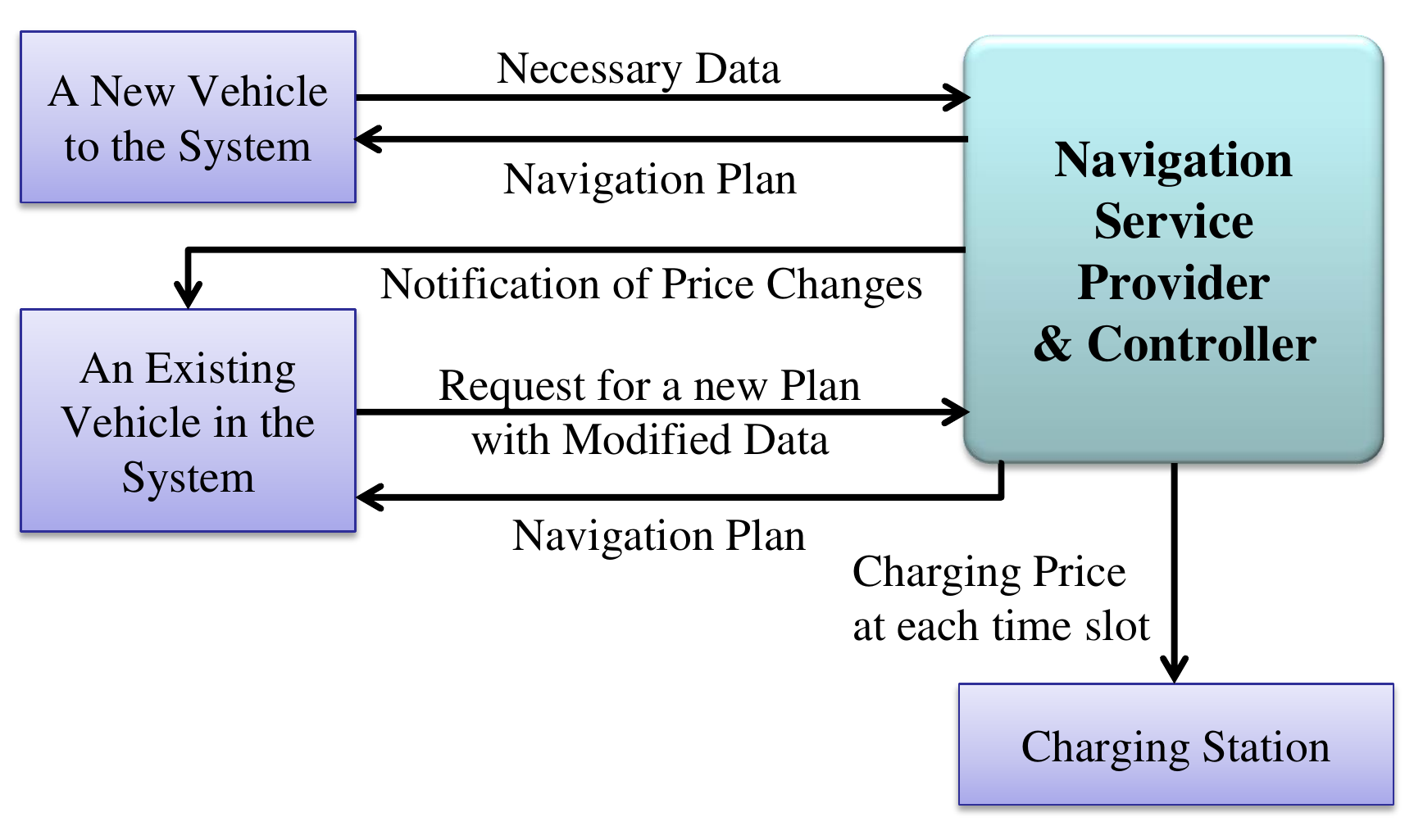}
%\vspace{-6pt}
\caption{The steps of the navigation control technique.}
\label{Fig_Control}
\end{center} 
%\vspace{-9pt}
\end{figure}
%%%%%%%%%%%%%%%%%%%%%% 

%\vspace{-6pt}
%%%%%%%%%%%%%%%%%%%%%%%%%%
\subsection{Overview of Navigation Control Technique}
\label{SSec:Idea_Control}

%Here, we propose a simple approach for traffic load balancing. The main idea here 
The basic idea we follow in our control technique is to adjust the charging prices of the stations at each time slot to indirectly distribute the vehicles in the roads and the charging stations. 
%% added by Qi
The control technique is shown in Figure~\ref{Fig_Control}, which works as follows:
%In the system, the service controller can communicate with the charging station and vehicles via wired or wireless network. 
%
When a PHEV  enters the road system, it connects to the controller with necessary information necessary to get the navigation route. The controller computes a satisfiable navigation plan (based on the model discussed in Section~\ref{SSec:SysModel}) and returns it to the vehicle. Hence, the controller has the navigation plans of all the existing PHEVs on the roads. The controller can follow a decentralized framework of computing systems to deal with an excessive number of PHEVs. With the time, the routing data about the PHEVs gets updated. The forecasted traffic (including non-PHEV vehicles) on the roads can also be changed due to updated situations.
%%With the updated these routing information, the controller can update the predicted loads 
%From these plans, 
Therefore, the controller have an updated information about the traffic on each road segment, and it adjusts the charging price (for a time slot) when the update is different than the estimation.
The price adjustment technique considers that the routing paths of the existing PHEVs are expected to remain unchanged. Hence, the technique considers the cost  requirement of each vehicle as a constraint, so that the navigation plan for the vehicle will remain valid even after any price adjustment.
%When a PHEV  enters the system, it communicates with the controller and sends necessary information to the controller. The controller computes a satisfiable navigation plan based on the model discussed in Section~\ref{SSec:SysModel} and returns it to the vehicle. 
%Hence, the controller has the navigation plans of all the existing vehicles in the system.  From these plans, the controller can know the traffic loads of each road segment and adjust the charging prices at each time slot. 
%The price adjustment technique considers that the existing routing paths are expected to be unchanged. Hence, the technique considers the cost  requirement of each vehicle as a constraint, so that the navigation plan for the vehicle will remain valid even after any price adjustment.
%
The controller can notify each existing vehicle in the system if there is a price adjustment. 
Then, a vehicle may request the controller for a new routing plan. 
%, may be along with a new set of requirements.
%The information flow in the system is shown in Figure~\ref{Fig_Control}.

In order to adjust the charging prices to achieve better load balance, 
the controller gives a weight for each point~$l$ associated to a charging station during a time slot $t$ based on the number of vehicles that have the point on its routing path during $t$. The busier the point is during $t$, the higher weight given to it at $t$. The control chooses the prices according to these weights$-$ higher weights get higher prices, while lower weights have lower prices.
The prices during a time slot ($t$) should be within some minimum ($P_{min,t}$) and maximum ($P_{max,t}$) bounds. 
%The more is the difference between the weights, the higher will be the difference between the prices. 
Moreover, the average of the charging prices of all the charging stations during a time slot $t$ is constrained to be equal to $P_{avg,t}$, where $P_{avg,t}$ is roughly equal to $(P_{min,t} + P_{max,t})/2$.

%\vspace{-6pt}
%%%%%%%%%%%%%%%%%%
\subsection{Navigation Control Model}
\label{SSec:Model_Control}

We define a list of parameters to model the navigation control technique. A solution to this model will provide the charging prices at the charging stations.

\vspace{3pt}
%%%%%%%%
%\subsubsection{Parameters for Modeling Navigation Control}
\noindent\textbf{Parameters for Modeling Navigation Control:}
\label{SSec:Param_Control}

We define $\mathbb{S}$ as the set of the charging stations, \ie the set of the points where the stations are located. We also define $\mathbb{V}$ as the set of the vehicles currently existing in the system. At the beginning of time slot $\bar{t}$, the controller updates the charging prices for the stations for a number of future time slots (\ie time $t \ge \bar{t}$) by modeling a constraint satisfaction problem. We use the following parameters in the model:

%\begin{itemize}
%\item 
\noindent\emph{Charging Cost}. $\bar{PE}_v$ denotes the total cost that the vehicle $v \in \mathbb{V}$ spent for charging the battery during the trip. On the way from the source to the destination the vehicle may need to charge its battery several times, here, $m$ times. We define $\bar{PE}_v$ as follows:
%\comment{Need to think more and then write on this.}
%$$\bar{PE}_v = Ps_{h_1,l_1,t_1} \times C_{h_1,l_1} + \cdots + Ps_{h_m,l_m,t_m} \times C_{h_m,l_m}$$
%
\begin{equation}\nonumber
\bar{PE}_v = (PE_{l_1})_v + \cdots + (PE_{l_m})_v
\end{equation}
Here, $l_i$ is the $i$'th point ($1 \le i \le n$) on the routing path from the source to the destination and $PE_{l_i}$ is already defined in Section~\ref{SSSec:Model_Time_Cost}. Remember that $PE_{l_i}$ depends on $Pe_{l_i}$, while $Pe_{l_i}$ depends on $Ps_{l_{\bar{i}},t_{\bar{i}}}$ ($\bar{i} \le i$), the charging price of the charging station located on point $l_{\bar{i}}$, where the vehicle charges its battery during time slot $t_{\bar{i}}$. In the case when time slot $t_{\bar{i}} < \bar{t}$, the vehicle has already charged its battery, while in the case when times slot $t_{\bar{i}} \ge \bar{t}$, the vehicle will charge its battery. Hence, the the updated prices have an impact on $\bar{PE}_v$ for the time slots equal or higher than $t$.
%There can be multiple paths (\ie multiple satisfiable solutions) from the source to the destination.

%\item 
\noindent\emph{Gasoline Cost}. $\bar{PG}_v$ denotes the total cost that the vehicle spent for gasoline during the trip. $\bar{PG}_v$ is computed as:
\begin{equation}\nonumber
\bar{PG}_v = (PG_{l_1})_v + \cdots + (PG_{l_n})_v
\end{equation}
Here, $l_i$ is the $i$'th point ($1 \le i \le n$) on the routing path from the source to the destination and $PG_{l_1}$ is already defined in Section~\ref{SSSec:Model_Time_Cost}. 
Since the gasoline price is kept unchanged, $\bar{PG}_v$ remains the same, irrespective of the change in the charging prices. 

%\item 
\noindent\emph{Weight}. $W_{l,t}$ denotes the weight of the vehicles passing through point ${h,l} \in \mathbb{S}$ during slot $t$ ($t \ge \bar{t}$), \ie the ratio of the number of vehicles passing (actually expected to pass according to the vehicles' navigation plans) through point~$l$ over the total number of vehicles passing all the points in the set $\mathbb{S}$ during $t$. 
$$W_{l,t} = \frac{Nv_{l,t}}{\sum_{l' \in \mathbb{S}}{Nv_{l',t}}}$$
Here, $Nv_{l,t}$ represents the total number of vehicles passing through point~$l$, which is computed from $X_{\hat{l}, l}$ for any $\hat{l}$ and $St_l$ associated with all vehicles. 
%\end{itemize}

\vspace{3pt}
%%%%%%%%
%\subsubsection{Modeling Navigation Control}
\noindent\textbf{Modeling Navigation Control:}
\label{SSec:Model_Control}

The gasoline cost $\bar{PG}_v$ for a vehicle and the weight $W_{l,t}$ for each gasoline station during a time slot are taken as computed value which are constant in the model. We synthesize the charging prices for each station at upcoming time slots (\ie for each time slot $t \ge \bar{t}$) while the charging prices during previous time slots are invariable. %\ie each time slot $t < \bar{t}$ is invariable. 
%We already discussed the idea of this synthesis process. The idea is modeled in the following.
%
The charging price of a gas station during a slot $t$ is proportional to the weights of the vehicles passing through the point associated to the station. This is formalized by the following relation:
\begin{equation}
\label{Eq_Relation}
(W_{l,t} > W_{l',t}) \rightarrow (Ps_{l,t} > Ps_{l',t})
\end{equation}

The prices cannot be less than $P_{min,t}$ and more than $P_{max,t}$ during time slot $t$. If the weight is zero, \ie no vehicles passing through the point corresponding to the charging station, the price is taken as the minimum. We consider the similar constraint for maximum weight (\ie 1). These constraints are formalized as follows:
\begin{equation}
(Ps_{l,t} \ge P_{min}) \wedge (Ps_{l,t} \le P_{max,t})
\end{equation}
\begin{equation}
(W_{l,t} = 0) \rightarrow (Ps_{l,t} = P_{min,t})
\end{equation}

It is also ensured that the average of the charging prices at different charging stations during time slot $t$ are equal to the (average) electricity price set for this time slot ($P_{avg,t}$) (Equation~(\ref{AvgPriceDecision})). This price can be identified based on the electricity price(s) set by the associated energy provider(s) and the profit margin set for the charging stations. 
\begin{equation}\label{AvgPriceDecision}
\frac{\sum_{h,l, S_l} {Ps_{l,t}}}{\sum_{l \in \mathbb{S}}{1}}  = P_{avg,t}.
\end{equation}

The crucial constraint is that the vehicles already in the system have to be able to reach their destination within the cost limit. Since the prices may change (for the upcoming time slots), we need to verify whether the cost for a vehicle $v$ remains within the limit $Cp_v$.
\begin{equation}
\label{Eq_Existing}
\bar{PG}_v + \bar{PE}_v \le Cp_v.
\end{equation}

%\vspace{-6pt}
%%%%%%%%%%%
\subsection{A Case Analysis}
\label{SSec:Case}

%%%%%%%%%%%%%%%%%%%%%%%
\begin{wrapfigure}{r}{0.5\textwidth}
\vspace{-24pt}
\begin{center}
\includegraphics[width=0.5\textwidth]{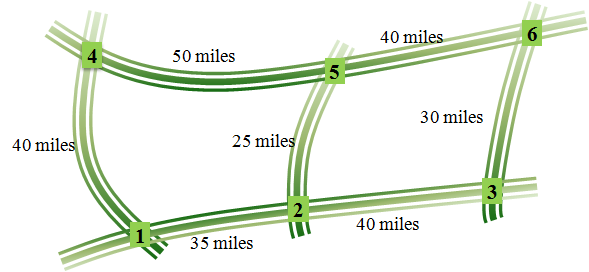}
\vspace{-12pt}
\caption{The topology of the road network used in the case analysis of Section~\ref{SSec:Case}.}
\label{Fig_Example_Control}
\end{center} 
\vspace{-12pt}
\end{wrapfigure}
\begin{figure}[t]
%\vspace{-12pt}
\begin{center}
        \subfigure[]{            
          \includegraphics[width=0.45\textwidth]{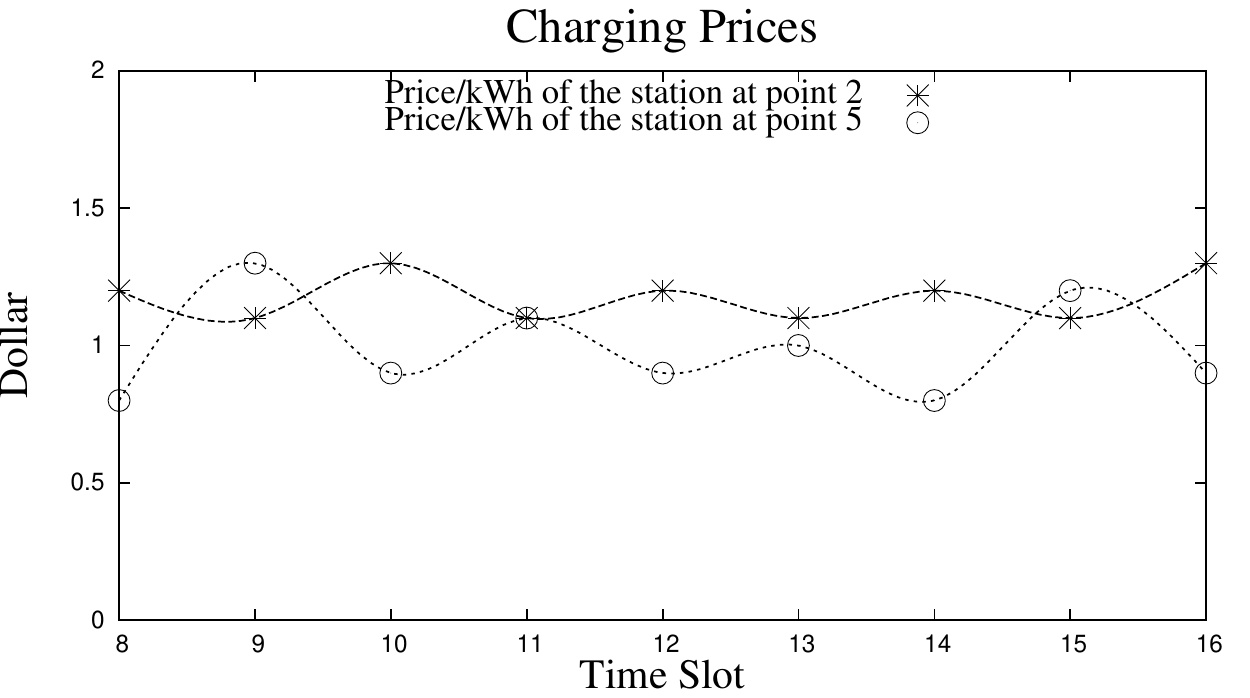} 
	  \label{Graph_Control_Analysis_A}
        }
        \subfigure[]{            
           \includegraphics[width=0.45\textwidth]{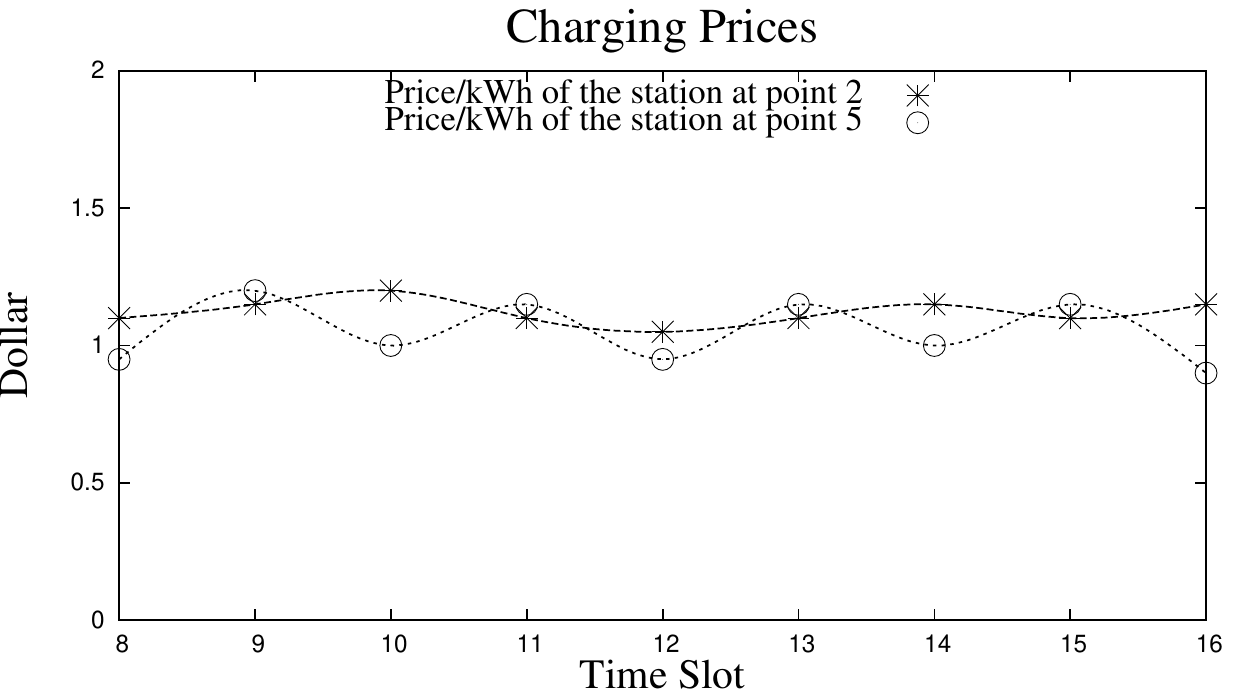} 
	   \label{Graph_Control_Analysis_B}
        }
\vspace{-9pt}
\caption{Change of charging prices with the time: (a) the gas stations have the same price rate, and (b) the gas stations have different price rates.}
\label{Graph_Control_Analysis}
\vspace{-6pt}
\end{center} 
\end{figure}

%%%%%%%%%%%%%%%%%%%%%% 
%
In Figure~\ref{Graph_Control_Analysis}, we show the change of the charging prices with the time slots. We consider the road network as shown in Figure~\ref{Fig_Example_Control}, where each location point has a charging station and a gas station. In this example, we take 100 vehicles, which are arbitrarily distributed during 24 time slots. For each vehicle, the source and destination, and the starting time of the trip are chosen randomly following uniform distribution. The maximum and minimum charging prices per kWh are taken as \$1.5 and \$0.5, respectively. The initial price for each charging station is taken randomly. In the first scenario, the gas price is fixed and the same for all gas stations (\$4.00/gallon) while in the second scenario, the gas stations have different prices. These prices are fixed throughout the day. The cost and time constraints are taken from the ranges of 10-50 dollars and 100-200 minutes, respectively. 

%%%%%%%%%%%%%%%%%%%%%%%%%
%%%%%%%%%%%%%%%%%%%%%%%%%

The results of the first case are presented in Figure~\ref{Graph_Control_Analysis_A}, which shows the charging prices for two stations: the charging stations located at point 2 and point 5. We see that the prices change with the time. It is interesting to observe that the charging price of the station at location 5 is most often less than the charging price of the station at point 2, except for a few cases. The reason behind this behavior is that the connecting roads to point 2 are shorter compared to the connecting roads to point 5. That is why the navigation management plans contain point 2 more than point 5, especially due to the longest road segment between 4 and 5. Hence, to divert the vehicles to point 5, most of the time the charging price of the station at point 2 is kept high.
The results associated with the second case are shown in Figure~\ref{Graph_Control_Analysis_B}. Here, we observe a different behavior in price changes because the gas prices of the stations on points 1, 2, and 3 are higher (\$4.00, \$3.90, and \$3.80, respectively) than that of the stations on points 4, 5, and 6  (\$3.90, \$3.80, and \$3.70, respectively). Therefore, the PHEVs have incentives to choose routes through points 4, 5, and 6 compared to the case in Figure~\ref{Graph_Control_Analysis_A}.

%\vspace{-3pt}
%%%%%%%%%%%%%%%%%%%%%%%%%%%%%%%%%%%%%%%%%%%%%%%%%%%
\section{Evaluation}%\vspace{-3pt}
\label{Sec:Evaluation}

%\comment{TODO... Discussion about the graphs}
We evaluate the scalability of our proposed navigation management model and control model. %We analyze the model using different synthetic highway systems. 
Due to the unavailability of the real-life data, we analyze the models using different synthetic data.
%We mainly present the evaluation results for the navigation management model, since we observed similar behaviors for the control model.

%\vspace{-6pt}
%%%%%%%%%%%%%
\subsection{Methodology}
\label{SSec:Methodology}

We evaluate the scalability of our proposed models by analyzing the time and memory required for synthesizing the outputs. %by varying mostly the problem size.
We apply efficient techniques for increased scalability of solving the navigation management model. %, particularly in terms of the time. 

%%%%%%%%
%\vspace{6pt}
\subsubsection{Evaluation Data Overview}

%In this revised version, we consider the real road system. We download the map from <map> in the <> format and preprocess the map data to our input format. We take the longitude and latitude of each location point, the points on a way/road, and speed limit of a road from the data. We calculate the distance between two points on a road using their coordinates. We consider primary and secondary highways only. We take points a minimum distance apart from each other, although the distance between two consecutive points on a road is calculated considering the coordinates of all the intermediate points.

The problem size depends mainly on the number of location points, \ie roads, as well as charging and gas stations. The problem size is primarily specified by the total number of location points. The location points are taken (or abstracted) based on the real road maps (particularly, motorways/highways in the Eastern United States). We download the map from OpenStreetMap~\cite{OpenStreetMap} in the OSM (Open Street Map) JSON file format and process the data to convert it to our input template. Since we consider long trips in this research, we download and process the roadways with the ``highway'' key values as ``motorway,'' ``trunk,'' and ``primary.''
%Although there is a shuttle difference between ``motorway'' and ``highway'' in the transport system, we use the term ``highway'' in general. 
We take the longitude and latitude of each location point, the points on a highway (using the ``ref'' and ``tiger/cfcc'' tags), and speed limit of a road from the data. The distance between two consecutive points on a highway is calculated considering the coordinates of the points. We consider points on a highway with a minimum distance apart from each other. The distances are kept within a range of 1$-$40 miles. The intersections/links between the roads are computed from the coordinates and the tags like ``motorway\_link'' and ``motorway\_junciton.'' 
%The number of locations points are considered up to 1000 for the evaluation. 
In the case of varying the problem size in terms of the number of links (or branch roads) while keeping the number of location points constant, we arbitrarily consider these links.

The charging stations at different locations are considered according to the Alternate Fuels Data Center, managed by the Office of Energy Efficiency and Renewable Energy, U.S. Department of Energy~\cite{DoE-EERE}. The gas station locations can be found from many sources. We get them OpenStreetMap (using the ``amenity'' tag)~\cite{OpenStreetMap}. However, we consider arbitrary locations for the charging or gas stations %in those experiments where 
when we vary their number to observe the impact of their availability %of stations 
on the scalability.
%proportionally with respect to the number of points. In an experiments, if otherwise not stated,  90\% of the points have gas stations, while 50\% have charging stations. 

%Due to the lack of real data, 
The rest of the evaluation data is synthetic.
%
%The number of stations is taken proportionally with respect to the number of points. In an experiments, if otherwise not stated,  90\% of the points have gas stations, while 50\% have charging stations. 
%The length of a road is taken arbitrarily from the range of 10$-$40 miles.
%In the synthetic road system, the number of locations points are taken up to 1000. %The routes are assumed to be parallel to each other and 
%A point on a route can be connected to a point on a neighboring route (as shown in Fig.~\ref{Fig_Example_Control}). 
The number of time slots in a day is considered as 24, where each time slot covers 60 minutes. %\comment{why not 24}
%The average vehicle speed on a road is randomly chosen between 0.5 to 1.5 miles/minute. 
For the gasoline and electricity prices, the ranges are taken from the ranges of \$3$-$\$5/gallon and \$0.5$-$\$1.5/kWh, respectively. %the range of \$3$-$\$5/gallon and the range of \$1$-$\$3/full charge. 
The capacity of a battery is taken randomly from the range of 10$-$20 KWh. We also randomly choose the mileage achieved by a vehicle from the range of 12$-$24 miles/gallon and the range of 8$-$16 miles/kWh. 
We encoded our model using Z3 .NET API and ran the verification of the model on an Intel Core i7 Processor with $16$ GB memory. %under Windows $10$ OS. 

%The problem size depends mainly on the number of location points, \ie roads, as well as charging and gas stations. The problem size is considered as the total number of location points. The number of stations is taken proportionally with respect to the number of points. In an experiments, if otherwise not stated,  90\% of the points have gas stations, while 50\% have charging stations. The length of a road is taken arbitrarily from the range of 10$-$40 miles. In the synthetic road system, 
%the number of locations points are taken up to 1000. %The routes are assumed to be parallel to each other and 
%A point on a route can be connected to a point on a neighboring route (as shown in Fig.~\ref{Fig_Example_Control}). The number of time slots in a day is considered as 24, where each time slot covers 60 minutes. %\comment{why not 24}
%The average vehicle speed on a road is randomly chosen between 0.5 to 1.5 miles/minute. For the gasoline and electricity prices, the ranges are taken from the ranges of \$3$-$\$5/gallon and \$0.5$-$\$1.5/kWh, respectively. %the range of \$3$-$\$5/gallon and the range of \$1$-$\$3/full charge. 
%The capacity of a battery is taken randomly from the range of 10$-$20 KWh. We also randomly choose the mileage achieved by a vehicle from the range of 12$-$24 miles/gallon and the range of 8$-$16 miles/kWh. 
%We encoded our model using Z3 .NET API and ran the verification of the model on an Intel Core i7 Processor with $16$ GB memory under Windows $7$ OS. 

%%%%%%%%%%%%
\begin{algorithm}[!t]
	\caption{Parallel Search for the Navigation Plan}
	\label{Algorithm_Parallel}
	
	\begin{algorithmic}[1]
		\Require $F$ := The navigation management model; \Comment{According to the given inputs}
		\Require $\mathbb{P}$ is the set of $n$ processes; \Comment{Can be threads}
		\For {Each Process $P \in \mathbb{P}$}
			\State $R_P$ := A random number;  \Comment{To randomize the navigation space exploration in $F$}
			\State Call $\mathit{find Navigation Plan}$($F_P$, $R_P$) of Process P;    			
  		\EndFor		
		
		\State $\mathit{Route}$ := NULL; 	\Comment{A navigation route is yet to receive.}
		\State $\mathit{Flag}$ := TRUE;	\Comment{Whether the loop should continue.}
		%\Comment{Whether the loop should continue to check if a process is successful.}
		\While {$\mathit{Flag}$}
			\State Wait for an arbitrary small period;
			%\State $U$ := $|\mathbb{P}|$;	\Comment{Number of processes that are still executing.}			
			\For {Each Process $P \in \mathbb{P}$}
				\If {$\mathit{Done}_P$ = TRUE} \Comment{Process $P$ completes its execution.}
					\State $\mathit{Flag}$ := FALSE;
					%\State $U$ := $U - 1$;	
					\If {$\mathit{Result}_P \neq$ NULL} \Comment{A SAT result is received and saved in $\mathit{Result}_P$.}
						%\State $\mathit{Flag}$ := FALSE;
						\State $\mathit{Route}$ := $\mathit{Result}_P$;
						\State Break;
					\EndIf
				\EndIf						
    			\EndFor
			
			\If {$\mathit{Flag}$ = FALSE}
				\For {Each process $P \in \mathbb{P}$}
					\If {$\mathit{Done}_P \neq$ TRUE}
						\State Terminate process $P$;
					\EndIf
	    			\EndFor		
			%\ELSIF { $U$ = 0}
				%\State $\mathit{Flag}$ := FALSE;			
			\EndIf
  		\EndWhile
	\end{algorithmic}
\end{algorithm}
%%%%%%%%%%%

%%%%%%%%
%\vspace{6pt}
\subsubsection{Efficient Execution Mechanism}

\paragraph{Parallel Execution of Multiple Instances} Since the navigation management model often needs to deal with numerous parameters and arithmetic constraints, finding an optimal navigation plan in a ``near real-time'' period is overwhelming for a large road network. Therefore, we devise a scalable mechanism for a controlled execution of several instances of the navigation management model parallelly so that a satisfiable navigation plan can be achieved within a practically acceptable time. 
Algorithm~\ref{Algorithm_Parallel} presents the mechanism that executes multiple processes simultaneously. All these processes explore the navigation space together. Each process runs the navigation management model with a random seed such that SMT searches in an arbitrary order (Lines~4 and 5 in Algorithm~\ref{Algorithm_Parallel}). As we look for one satisfiable navigation route, the parent process keeps checking if a process ends (Lines~8 and~9). A process is successful at finding a path when it receives a SAT result (Lines~12). When the result is UNSAT, there is no navigation route. When a navigation route is found, the rest of the processes (which are yet to be completed) are terminated since we look for a single navigation route. On the other hand, if a process (\ie the model instance running by this process) fails to find a path, there is no way other processes can be successful, as the solved model has already explored the whole search space. Therefore, the rest of the processes are terminated (Lines~18-~23). 
%We can also divide the search in such a fashion that although each process explores a distinct search space, they collectively explore the whole space. In this case, the model execution will be faster as the search space is reduced according to the initial exploration direction (that make the search distinct). \comment{A proof is needed}

\paragraph{Search Space Pruning} The efficiency of the navigation route synthesis can be further improved by pruning the search space with respect to the source and destination. The idea is explained here. The navigation route depends on the source and the destination, along with the road network. Although the given road network can be large, according to the source and the destination there are often many points inc the network, which will not practically be advantageous to traverse to reach the destination. Those points can easily be pruned from the search space by initializing corresponding $X_l$s as false. Such a constraint can be a user's preference. For example, a simple pruning condition can be as follows. If the summation of the shortest distance from the starting point to a point~$l$ (let it be $\hat{D}_{\mathit{Sv}, l}$) and that from the point to the destination point ($\hat{D}_{\mathit{Sv}, l}$) is larger than $r$ times (\eg twice) the shortest distance between the source and the destination ($\hat{D}_{\mathit{Sv}, \mathit{Dv}}$), the point should not be traversed to reach the destination. 
%(\ie the Pythagorean distance considering the geographical coordinates of each point). 
The corresponding formalization is presented below:
\begin{equation}\nonumber
%(\sqrt(x_l - x_{S_v})^2  + (x_{ D_v} - x_l)^2  >= r \times (D_v - S_v)^2) ~ \rightarrow \neg X_l
(\hat{D}_{\mathit{Sv}, l} + \hat{D}_{l, \mathit{Dv}})  \ge r \times \hat{D}_{\mathit{Sv}, \mathit{Dv}} ~ \rightarrow ~\neg X_l
\end{equation}

Since the shortest distances are expected to be already known, the time for space pruning is insignificant. 

%%%%%%%%
\subsubsection{Evaluation Approach}

We evaluate the scalability of  the proposed navigation management models on various scenarios in terms of the problem size as well as the navigation requirements. 
We mainly present the evaluation results for the navigation management model, since we observe similar behaviors for the control model.
We do not evaluate the model by performing a comparison with a related work because to the best of our knowledge there is no existing work that considers the same problem like ours. We consider a comprehensive list of routing properties for navigation planning. GreenGPS~\cite{Raghu11, Saremi16}, a closest work to this research, provides fuel-efficient routes, considering stop lights and the traffic congestion. The technique does not consider recharging (or refueling) requirements. In highways where there are usually no stop light or traffic congestion, this tool performs like traditional GPS services.

%\vspace{-6pt}
%%%%%%%%%%%%%%%
\subsection{Evaluation Results}
\label{SSec:Results}

%%%%%%%%%%%%%%%%%%%%%%%%
\begin{figure*}[t] 
%\vspace{-12pt}   
    \begin{center}
        \subfigure[]{
            \includegraphics[width=0.45\textwidth]{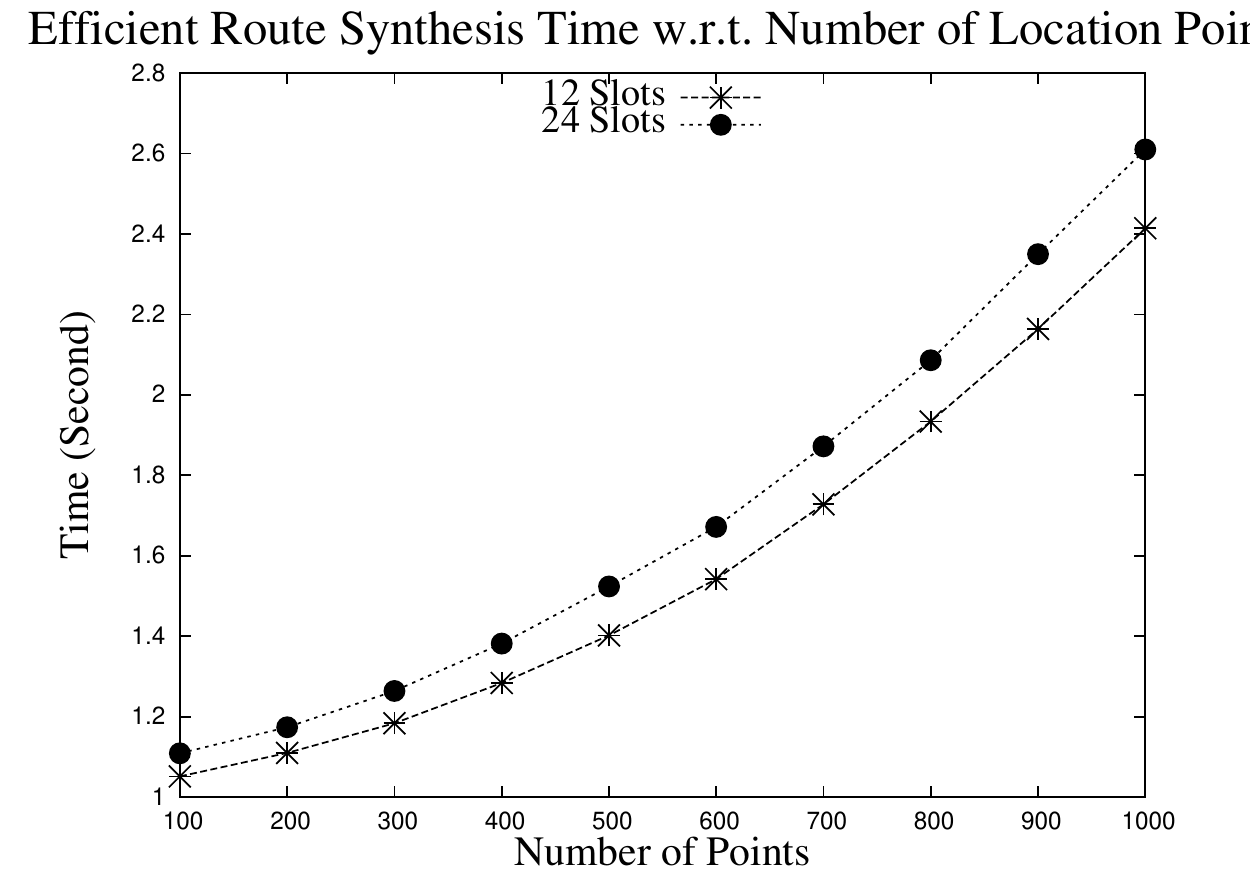}
            \label{Graph_Time_Points}
        }
	\subfigure[]{
            \includegraphics[width=0.45\textwidth]{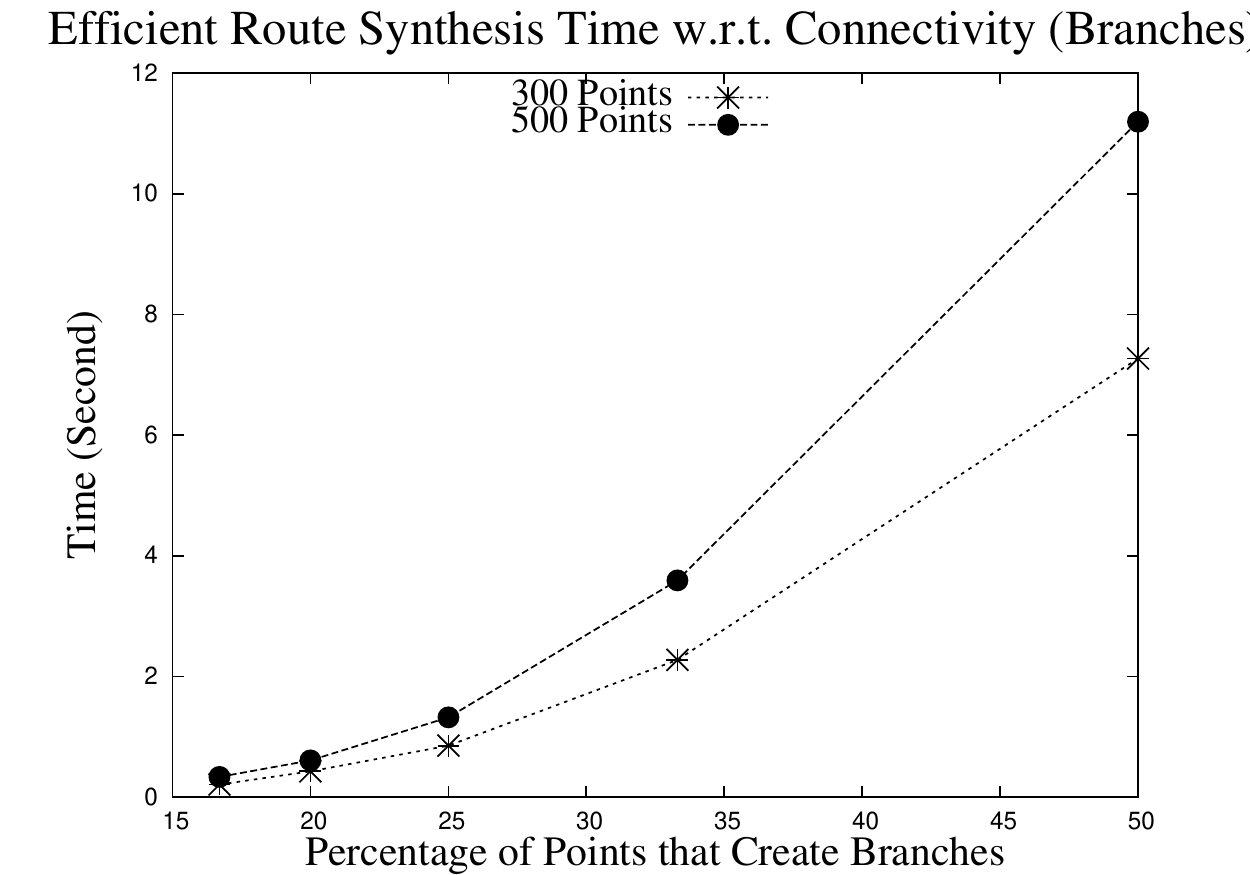}
            \label{Graph_Time_Links}
        }
%         \subfigure[]{
%           \label{Graph_Time_Stations}
%           \includegraphics[scale=0.45,keepaspectratio=true]{Graphs/G_Time_Stations.pdf} 
%        }
%	\subfigure[]{
%           \label{Graph_Time_Distance}
%           \includegraphics[scale=0.45,keepaspectratio=true]{Graphs/G_Time_Distance.pdf} 
%        }
    \end{center}
    \vspace{-9pt} 
    \caption{The navigation management plan synthesis time with respect to: (a) the number of points and (b) the number of links (particularly, the number of points connected with the neighboring route).}
%    \caption{The analysis time, \ie the navigation management plan synthesis time, with respect to: (a) the number of points, (b) the number of links (particularly, the number of points connected with the neighboring route), (c) the number of stations, and (d) the trip distance.}
%\vspace{-6pt}    
\end{figure*}
%%%%%%%%%%%%%%%%%%%%%%%%

%%%%%%%%%%%%%%%%%%%%%%%%
\begin{figure*}[t] 
%\vspace{-12pt}   
    \begin{center}
        \subfigure[]{
            \includegraphics[width=0.45\textwidth]{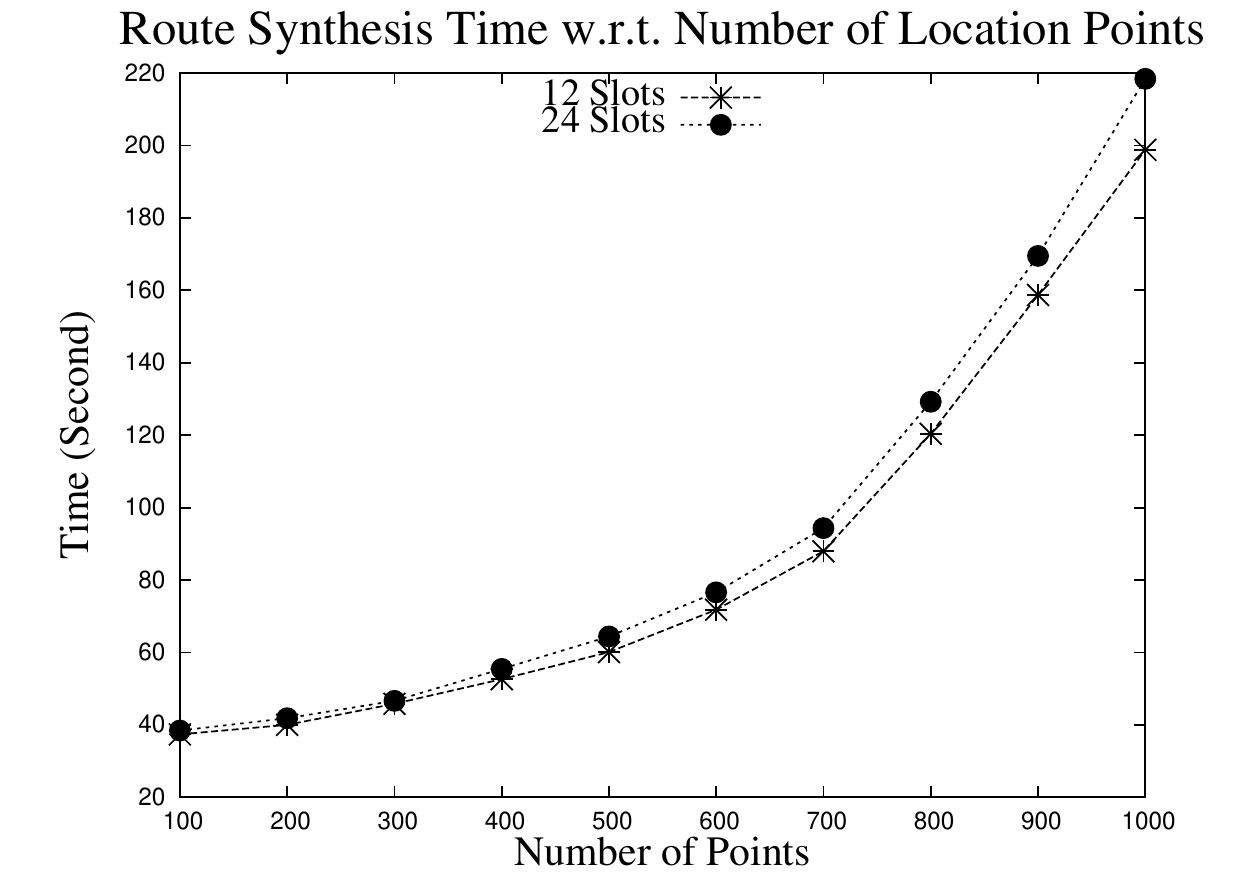}
            \label{Graph_Time_Points2}
        }
	\subfigure[]{
            \includegraphics[width=0.45\textwidth]{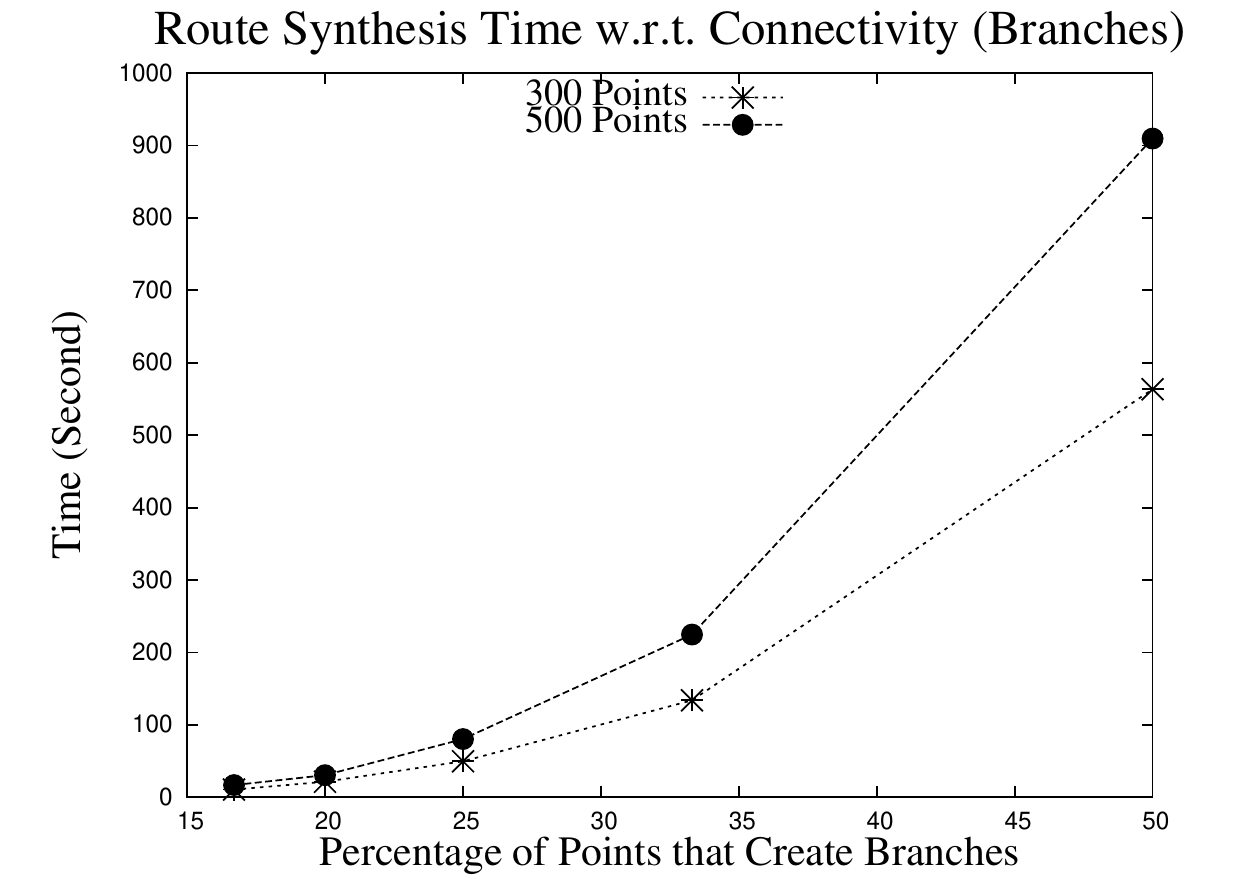}
            \label{Graph_Time_Links2}
        }
%         \subfigure[]{
%           \label{Graph_Time_Stations}
%           \includegraphics[scale=0.45,keepaspectratio=true]{Graphs/G_Time_Stations.pdf} 
%        }
%	\subfigure[]{
%           \label{Graph_Time_Distance}
%           \includegraphics[scale=0.45,keepaspectratio=true]{Graphs/G_Time_Distance.pdf} 
%        }
    \end{center}
    \vspace{-9pt} 
    \caption{The navigation management plan synthesis time, when neither parallelism nor space pruning mechanism is applied, with respect to: (a) the number of points and (b) the number of links (particularly, the number of points connected with the neighboring route).}
%    \caption{The analysis time, \ie the navigation management plan synthesis time, with respect to: (a) the number of points, (b) the number of links (particularly, the number of points connected with the neighboring route), (c) the number of stations, and (d) the trip distance.}
%\vspace{-9pt}    
\end{figure*}
%%%%%%%%%%%%%%%%%%%%%%%%

\noindent
\textbf{Impact of the Problem Size:}
Figure~\ref{Graph_Time_Points} shows the model verification time, \ie the navigation management plan synthesis time with respect to the problem size. 
%In this analysis, we consider that 25\% of the location points are connected with a neighboring point, 50\% have charging stations, and 90\% have gas stations. We vary the problem size by changing the number of routes and the number of locations on each route. 
As shown in Figure~\ref{Graph_Time_Points}, we observe that the analysis time is super-linear with respect to the number of location points.
We consider two scenarios in this analysis. In one scenario the number of time-slots is 12, while in another scenario, this number is 24. We also observe that when the number of time-slots is larger, it takes a longer time. The increase in the problem size expands the model space, \ie the search space for the potential solution. As a result, we observe greater synthesis time for larger problems. 
Figure~\ref{Graph_Time_Links} shows the impact of the number of connecting roads on the navigation plan synthesis time. Unlike the experiment regarding Figure~\ref{Graph_Time_Points}, we vary the percentage of location points that are connected with a neighboring point (\ie the branch roads). We consider 24 time slots and perform experiments in two scenarios $-$ one with 300 points and another with 500 points. As we see in Figure~\ref{Graph_Time_Links}, the analysis time increases rapidly when the number of branch roads increases. Due to more connecting roads, there are many options to reach a specific location, which in turn increases the search space. 
It is worth mentioning that the devised efficient mechanism of executing the synthesis model (\ie the parallelism and the pruning techniques) significantly improves the scalability. Figures~\ref{Graph_Time_Points2} and~\ref{Graph_Time_Links2} show the navigation path synthesis time when neither parallelism nor pruning is applied. The results demonstrate the obvious improvement, when we compare them with that in Figures~\ref{Graph_Time_Points} and~\ref{Graph_Time_Links}. 

\begin{figure*}[t] 
%\vspace{-12pt}   
    \begin{center}
%        \subfigure[]{
%            \includegraphics[scale=0.45,keepaspectratio=true]{Graphs/G_Time_Points.pdf}
%            \label{Graph_Time_Points}
%        }
%	\subfigure[]{
%            \includegraphics[scale=0.45,keepaspectratio=true]{Graphs/G_Time_Links.pdf}
%            \label{Graph_Time_Links}
%        }
         \subfigure[]{
           \label{Graph_Time_Stations}
           \includegraphics[width=0.45\textwidth]{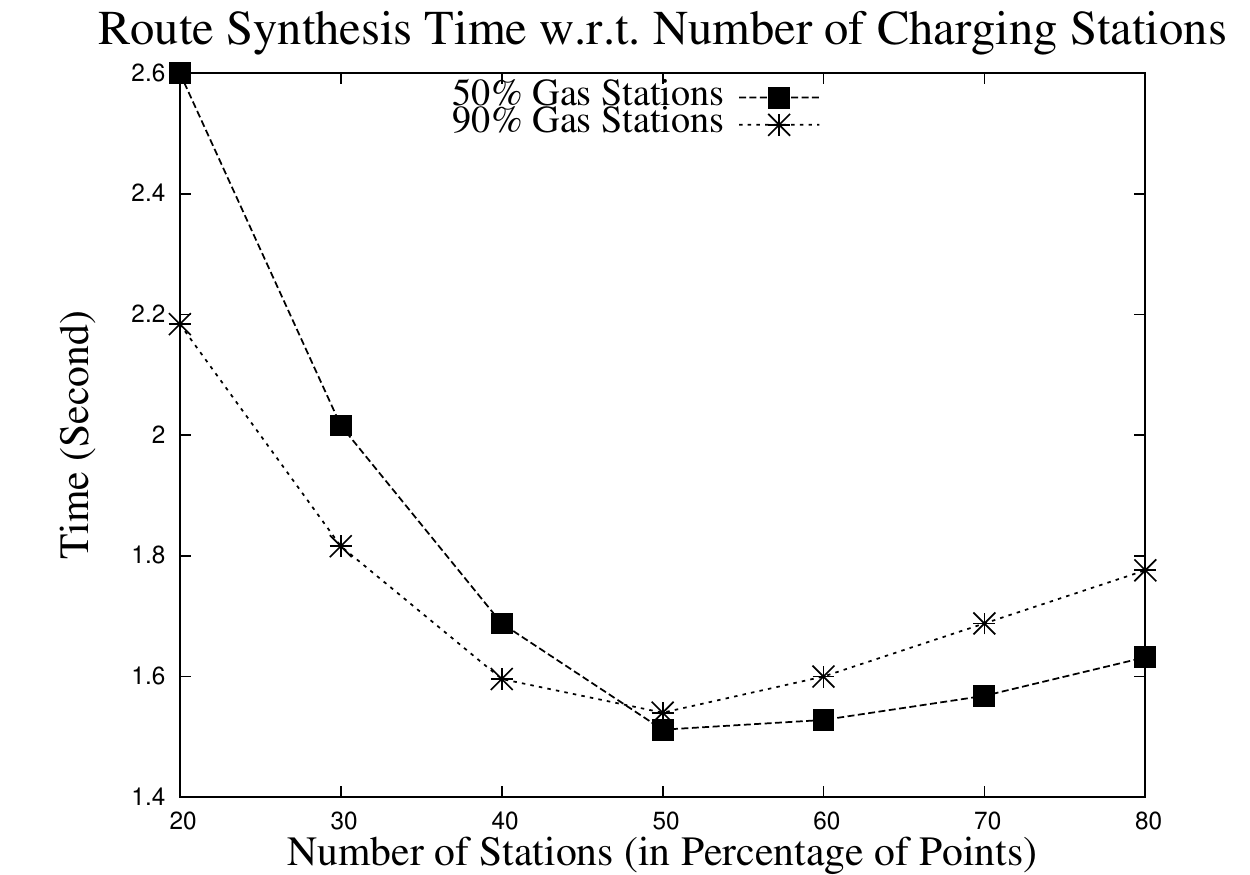} 
        }
	\subfigure[]{
           \label{Graph_Time_Distance}
           \includegraphics[width=0.45\textwidth]{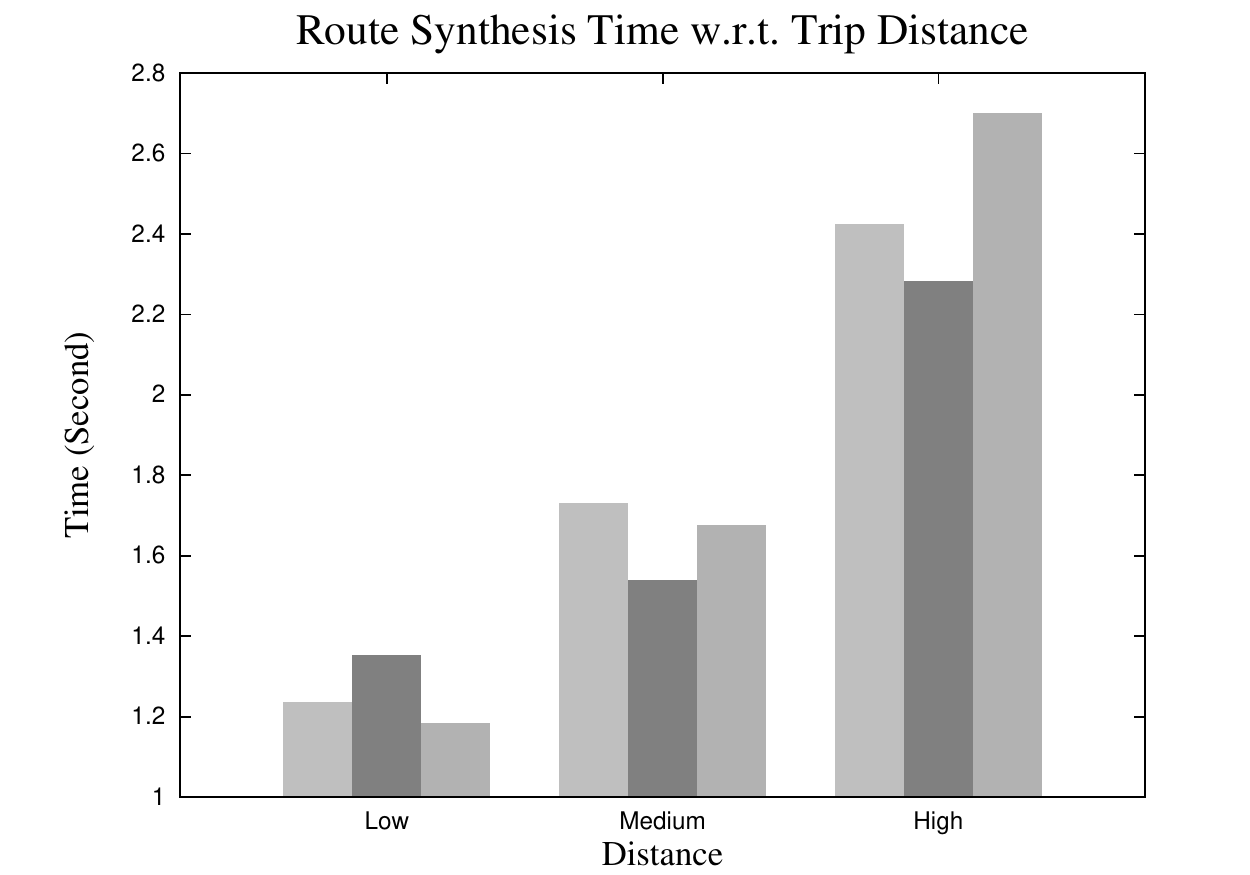} 
        }
    \end{center}
    \vspace{-9pt} 
    \caption{The navigation management plan synthesis time with respect to: (a) the number of stations and (b) the trip distance.}
%    \caption{The analysis time, \ie the navigation management plan synthesis time, with respect to: (a) the number of points, (b) the number of links (particularly, the number of points connected with the neighboring route), (c) the number of stations, and (d) the trip distance.}
%\vspace{-6pt}    
\end{figure*}

The cost incurred in a trip depends on the distance driven and the type of fuel used by the vehicle. The more electricity (stored in the battery) is used as fuel instead of gasoline, the less the cost. As we know a battery requires frequent recharging, the availability of the charging stations is important. 
%The vehicle may also need refueling to reduce the time requirement in the case of recharging. 
However, if there is high availability of charging stations, there are many alternatives available for recharging the battery. If there are too many alternatives, the search space becomes larger and the synthesis time increases accordingly. %many of which does not lead to the satisfiable solutions. 
On the other hand, if the number of charging stations becomes very small (\eg less than 40-50\% of the points have charging stations), there are fewer alternatives for satisfiable solutions. As a result, the solver needs to cover a larger space to find a solution. Figure~\ref{Graph_Time_Stations} justifies this argument. 
A vehicle may also need refueling to reduce the time requirement $-$  availability of charging stations is less while battery recharging takes a longer time). Therefore, there is an impact from the availability of gas stations as well, which is reflected in  Figure~\ref{Graph_Time_Stations} as well. In this analysis, we consider 500 location points and two scenarios, one with gas stations at 50\% of the points while another has them at 90\% of the points. 
%It shows that, if the ratio (in percentage) increases (from 60\% to 100\%), the analysis time decreases. However, recharging takes a significant amount of time, which increases the amount of time spent for traveling. As a result, very often it is not possible to choose many stations for recharging, as the time spent for traveling has to satisfy the time constraint. 
%\comment{Varies the number of processes/threads to see the impact on time.}

%%%%%%%%%%%%%%%%%%%%%%%%
%%%%%%%%%%%%%%%%%%%%%%%%

%%%%%%%%%%%%%%%%%%%%%%%%
\begin{figure*}[t] 
%\vspace{-9pt}   
    \begin{center}
   	\subfigure[]{
            \label{Graph_Time_Via}
           \includegraphics[width=0.45\textwidth]{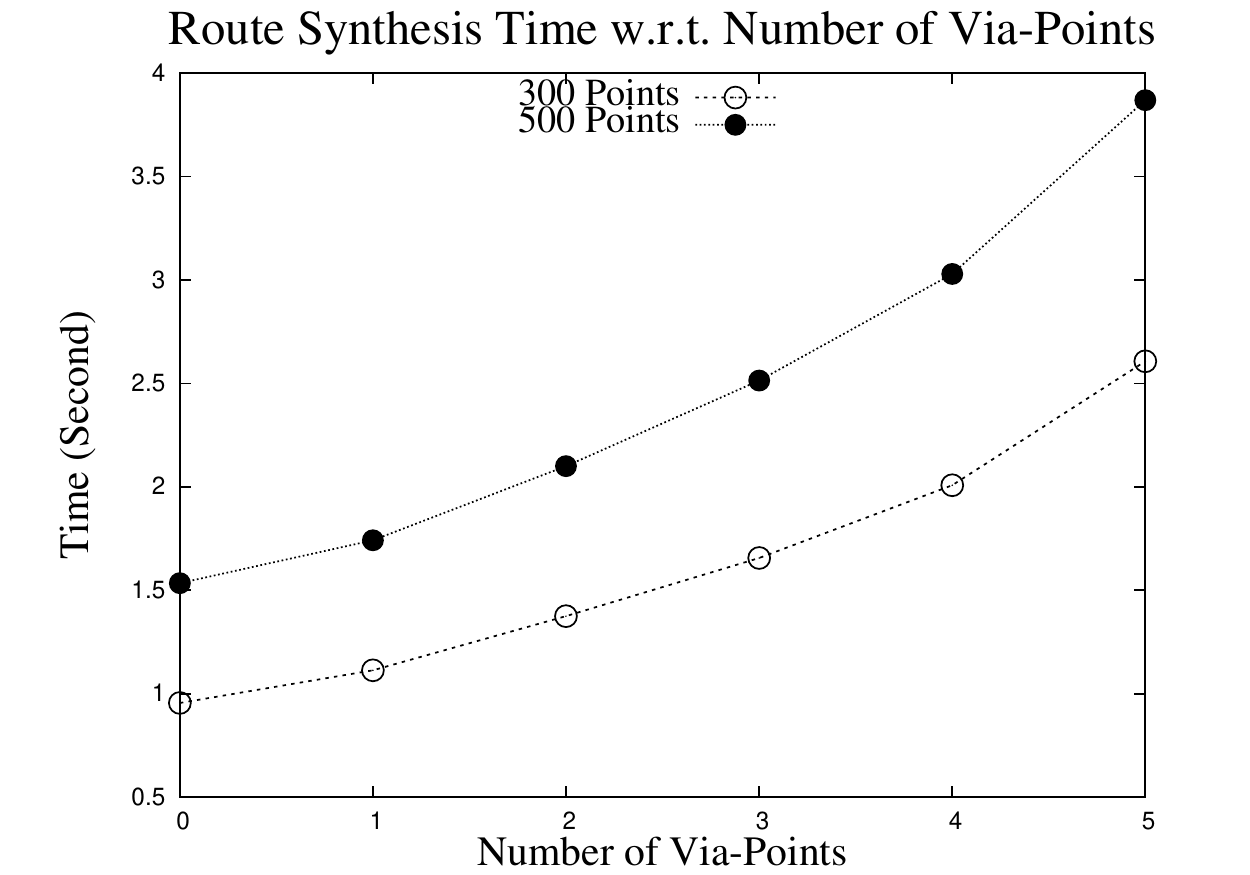} 
        }                           
	\subfigure[]{
            \label{Graph_Time_Control}
           \includegraphics[width=0.45\textwidth]{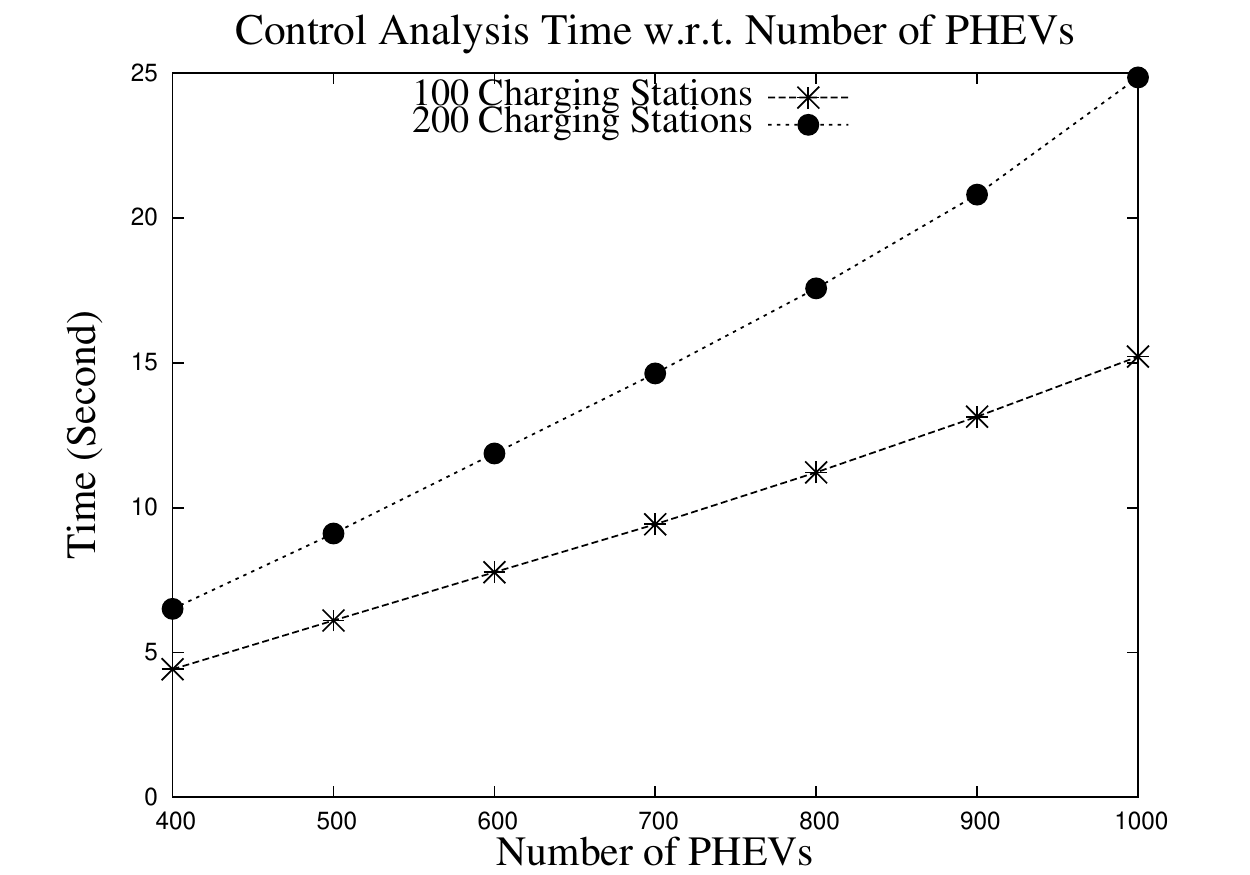}
        }                           
    \end{center}
    \vspace{-9pt} 
    \caption{The analysis time, \ie the navigation management plan synthesis time, with respect to: (a) the via-points, and (b) the impact of the number of vehicles on executing the navigation control model.}
%\vspace{-9pt}    
\end{figure*}
%%%%%%%%%%%%%%%%%%%%%%%% 

\noindent
\textbf{Impact of the Trip Distance:}
The distance (expressed in terms of location points) between the source and the destination has an impact on the navigation synthesis time. The analysis result is shown in Figure~\ref{Graph_Time_Distance} with respect to three qualitative distances: low (10-20 points), medium (50-60 points), and high (90-100 points), in a road system of 500 location points. We take three different arbitrary scenarios (in terms of source and destination) for each qualitative distance. The farther the destination from the source, %in order to reach the destination, \ie satisfy the constraints, 
the more alternatives there are to reach the destination satisfying the constraints. These alternatives mean navigation routes and recharging/refueling choices. As a result, the synthesis time increases with the increase in the trip distance.

\begin{figure*}[t] 
%\vspace{-12pt}   
    \begin{center}
        \subfigure[]{
           \label{Graph_Constraint_Time}
           \includegraphics[width=0.45\textwidth]{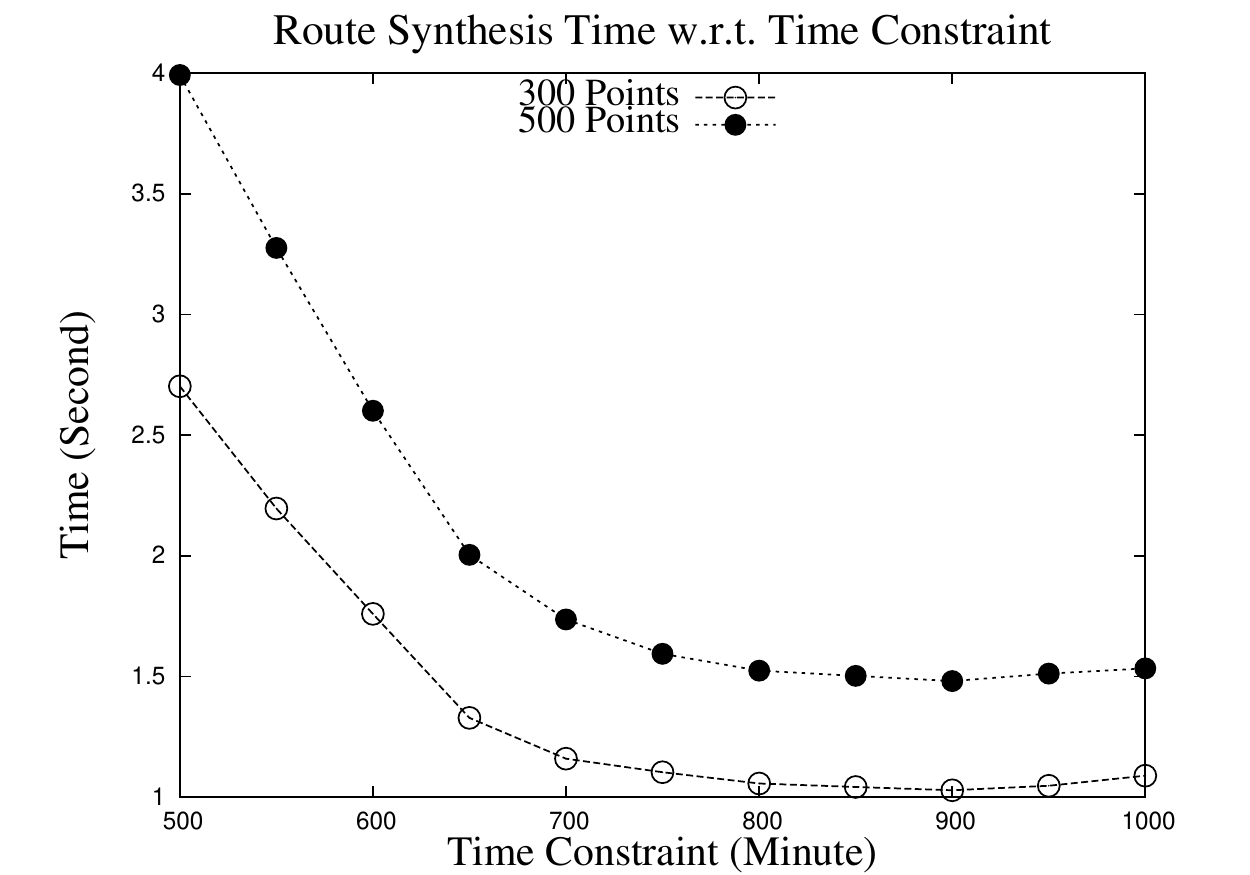} 
        }
         \subfigure[]{
           \label{Graph_Constraint_Cost}
           \includegraphics[width=0.45\textwidth]{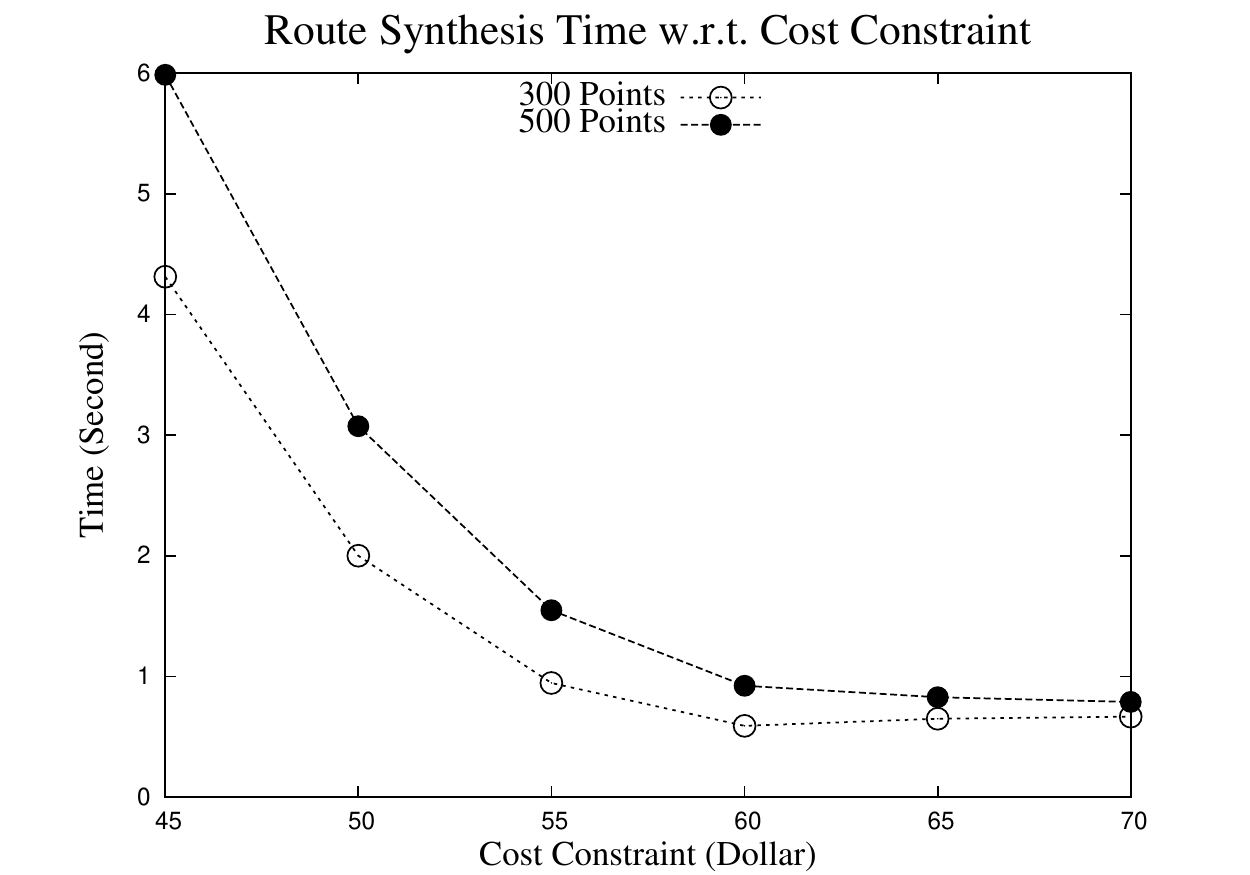} 
        }
%	\subfigure[]{
%           \label{Graph_Unsat}
%           \includegraphics[scale=0.45,keepaspectratio=true]{Graphs/G_Sat_Unsat.pdf} 
%        }                           
    \end{center}
    \vspace{-9pt} 
\caption{The impact of constraints on the navigation management plan synthesis time: (a) the time constraint  and (b) the cost constraint.}
%    \caption{(a) The impact of time constraint on the navigation management plan synthesis time, (b) the impact of the cost constraint on the navigation management plan synthesis time, and (c) the analysis time in the unsatisfiable cases.}
\label{Graph_Constraint}
%\vspace{-9pt}    
\end{figure*}

\noindent
\textbf{Impact of the Points of Interest:}
We observe that the number of intermediate points of interest (\ie via-points) impacts the synthesis time. The results are shown in Figure~\ref{Graph_Time_Via} for two different problem sizes. The figure shows that time increases with the increase in the number of via-points. Since the number of intermediate destinations increases and each destination is associated with a time constraint to be reached, a larger time is required to find a solution. However, the execution time is also dependent on how tight time and cost constraints are associated with these via points and the ultimate destination. This impact is illustrated later in Figure~\ref{Graph_Constraint}.  %As a result, the analysis time increases.

%%%%%%%%%%%%%%%%%%%%%%%%
%%%%%%%%%%%%%%%%%%%%%%%%

\noindent
\textbf{Impact of the Problem Size on Control Mechanism Synthesis:}
We also evaluate the scalability of our control model, and the results are presented in  Figure~\ref{Graph_Time_Control}, which shows the synthesis time with respect to the number of vehicles. In these experiments, we consider 500 location points for two different numbers of charging stations (100 and 200). We observe that the time for synthesizing the charging prices increases almost linearly with the increase in the number of vehicles.

\noindent
\textbf{Impact of the Constraint Tightness:}
We analyze the impact of the tight/ relaxed constraints on the model synthesis time. Tightening (relaxing) a time/cost constraint means decreasing (increasing) the time/cost constraint value. The analysis result is shown in Figure~\ref{Graph_Constraint_Time} where we vary the time constraint value. 
%and consider two scenarios sizes, 300 points and 500 points (\ie the number of routes is 5 and the number of points in each route is 50). 
We observe that the execution time increases with the reduction of the time constraint value because tightening a constraint reduces the number of possible solutions to the model; as a result, more searches are usually required to find a solution. We also observe the same behavior in the case of the cost constraint (Figure~\ref{Graph_Constraint_Cost}).

%%%%%%%%%%%%%%%%%%%%%%%%
\begin{figure*}[t] 
%\vspace{-12pt}   
    \begin{center}
%        \subfigure[]{
%           \label{Graph_Constraint_Time}
%           \includegraphics[scale=0.45,keepaspectratio=true]{Graphs/G_Time_Time.pdf} 
%        }
%         \subfigure[]{
%           \label{Graph_Constraint_Cost}
%           \includegraphics[scale=0.45,keepaspectratio=true]{Graphs/G_Time_Cost.pdf} 
%        }
	\subfigure[]{
           \label{Graph_Process}
           \includegraphics[width=0.45\textwidth]{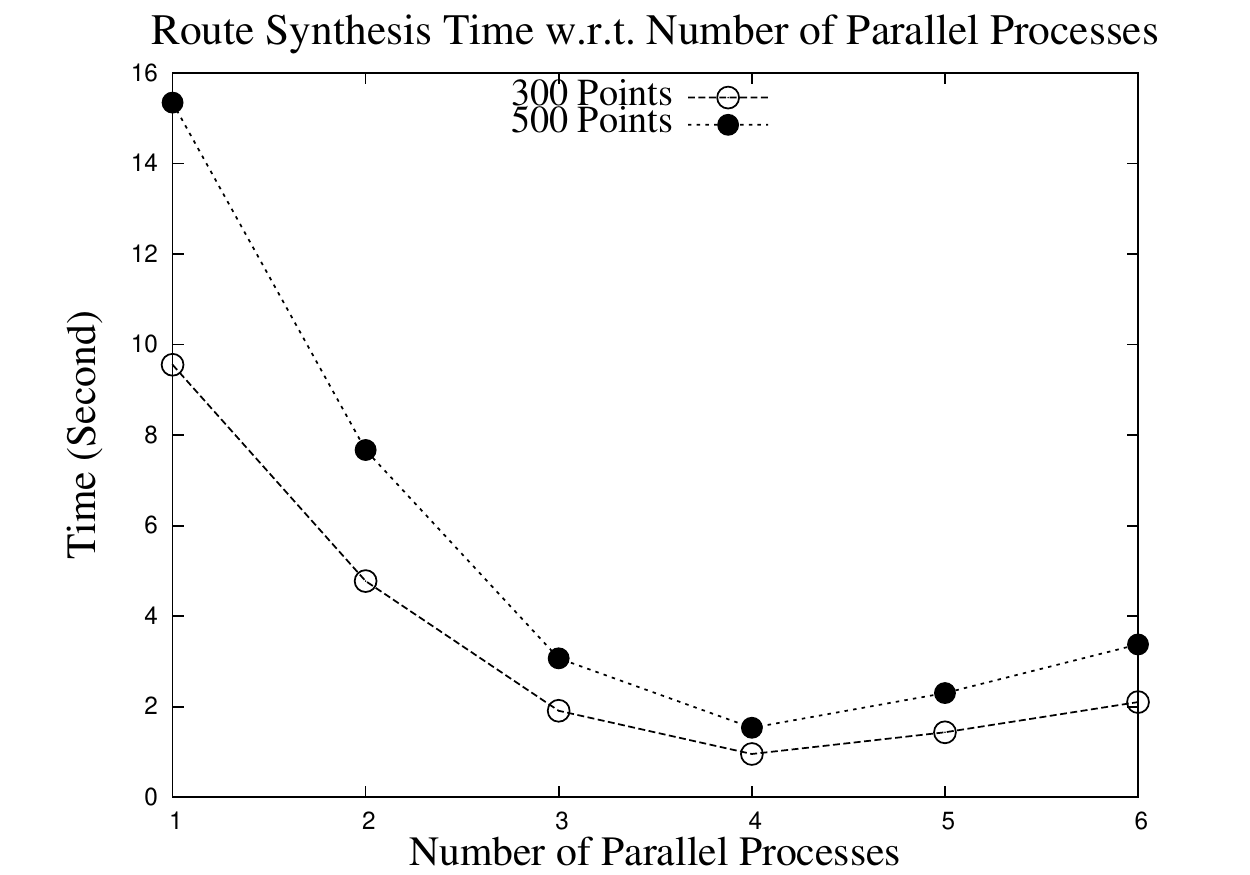} 
        }                           
	\subfigure[]{
           \label{Graph_Memory}
           \includegraphics[width=0.45\textwidth]{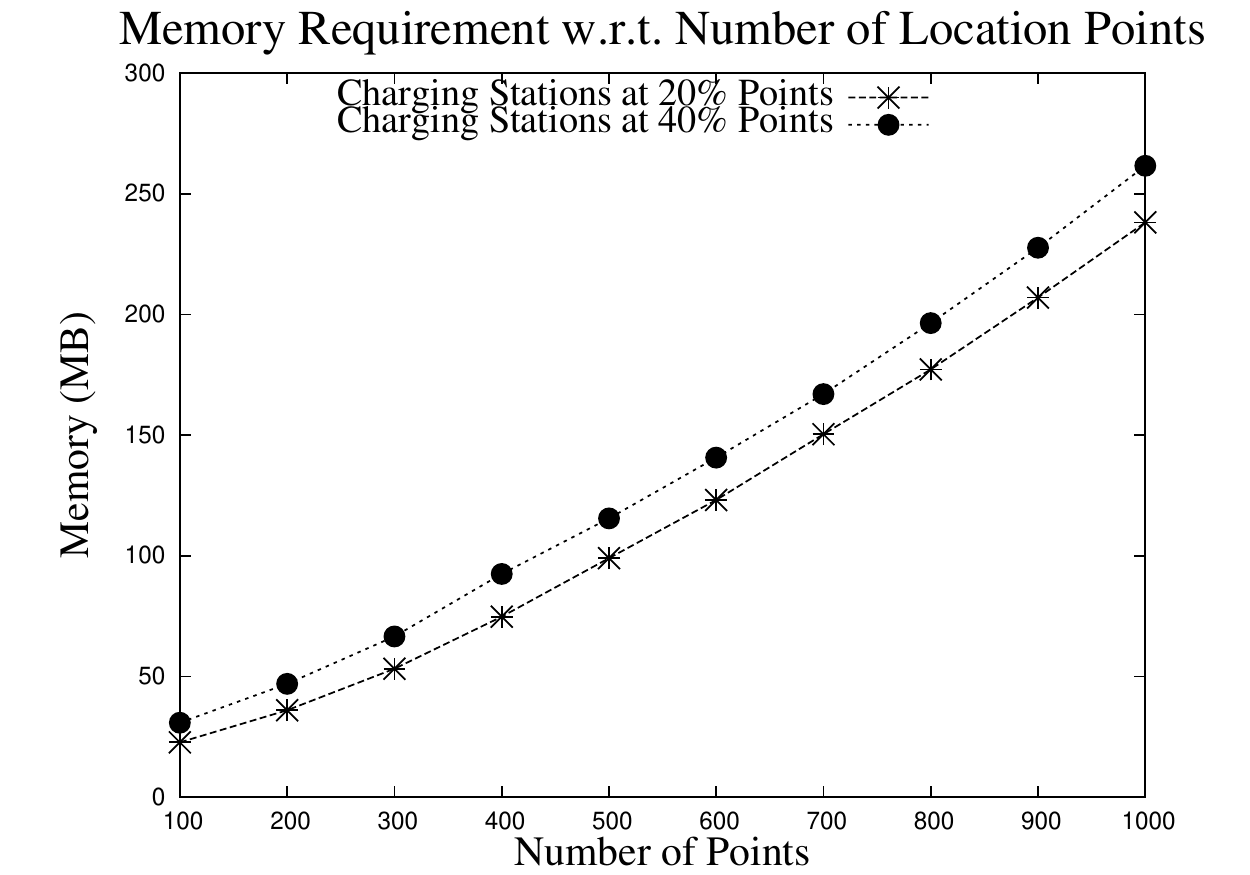} 
        }   
    \end{center}
    \vspace{-9pt} 
    \caption{(a) The navigation route synthesis time with respect to the number of processes executing the synthesis model parallelly and (b) the memory requirement with respect to the number of points.}
%\vspace{-9pt}    
\end{figure*}
%%%%%%%%%%%%%%%%%%%%%%%%

\noindent         
\textbf{Impact of Number of Parallel Processes:} 
In the above mentioned analyses, four processes are executed parallelly to synthesize the navigation route according to Algorithm~\ref{Algorithm_Parallel}. Figure~\ref{Graph_Process} presents the impact of the number of processes on the generation time. As the figure shows, the more is the number of processes, the smaller the execution time. Since we run the processes on a single physical host, the processes need to share the resources, and thus the overhead goes up when the number of sharing processes is high as well as the problem size, compared to the number of processing cores of the host (which is four in our case). Hence, there is a trade-off between the efficiency and the overhead. 

\noindent
\textbf{Memory Complexity:}
The memory requirement of the SMT solver~\cite{Z3} for our model is evaluated by changing the number of points. %(\ie the changing the highways, while keeping the number of exits per highway the same). 
The evaluation is done considering the memory required for encoding the problem in two scenarios with respect to the charging stations. In scenario 1, 20\% location points have charging stations, while in scenario 2, they exist at 40\% points. %The gas stations are considered at 90\% of the points.
The required memory for a model synthesis is the sum of the memory for modeling the system properties and the memory for modeling the constraints.
%The analysis result is shown in Table \ref{tab_space}. 
The analysis result is shown in Figure~\ref{Graph_Memory}. We observed that the memory requirement increases linearly with the increase in the number of location points. 
%However, the corresponding analysis/verification time, as shown in the table, increases exponentially.
%If the model size increases significantly, the SMT solver fails to give a solution. 
An increase in the model size depends on the problem size, mainly the number of location points and that of the charging and gas stations. 

%\balance

%%%%%%%%%%%%%%%%%%%%%%%%%%%%%%%%%%%%%%%%%%%%%%%%%%%
\section{Discussion}
\label{Sec:Related}

Here we discuss some concerns or aspects of the proposed navigation management solution.

%%%%%%%%%%%%%%
\subsection{Optimal Solution}

We model the navigation planning logically as a constraint satisfaction problem and solve using an efficient SMT solver. 
%The model does not abstract the requirements, except follow an input template for the road system, which is flexible for different input structures. 
The solver provides a solution if there is one satisfying the constraints on the given road system. The solution may not be the optimal one. However, the model can provide all possible routes to the destination, including the optimal one(s), if it is solved for all solutions (\ie complete exploration of the solution space). The tighter are the constraints (\eg the time and cost to reach the destination constraints), the smaller the solution space (\ie the number of alternative solutions) and the closer a solution to the optimal. The constraints can be too tight to get a solution (\ie no feasible route). However, a smaller solution space often takes a larger execution time. The impacts of time and cost constraints on the running time are shown in Figures~\ref{Graph_Constraint_Time} and~\ref{Graph_Constraint_Cost}. The tight constraints usually lead into larger execution times.

%%%%%%%%%%%%
\subsection{Usability}
%We apply some mechanisms, such as parallel execution and routing path pruning, to reduce the computation time. Within our limited computing capacity, our evaluation results show a few seconds of execution time for a large problem space. We present the impact of the number of parallel processes in Figure 13(a). In real practice, particularly in runtime planning cases, the computing capacity (i.e., using cloud/edge computing resources) can easily be extended to further reduce the execution time to an expected level.

The navigation route synthesis needs to be done within a reasonable time, which can be expected to be in run-time. %, so that the tool is usable in practice. 
Although the execution of the formal model follows an exponential scale of growth, the applications of parallel execution and routing path pruning, as we have presented in Section~\ref{SSec:Methodology}, significantly reduce the computation time.
%Our evaluation results have shown that the synthesis takes a few seconds. 
Within our limited computing capacity (i.e., limited processing power and memory), our evaluation results show a few seconds of execution time for a large problem space. We present the impact of the number of parallel processes in Figure~ \ref{Graph_Process}. In practical uses, particularly for run-time planning of the navigation paths, the computing capacity can easily be extended to further reduce the execution time.
%to an expected level.

Since parallelism is used, a powerful multi-core processor is required to get the desired performance. It can be argued that a PHEV is unexpected to be equipped with an IoT device (deployed in the vehicle or the user's smart device) which has sufficient processing ability for such a synthesis.
%Nowadays there are a lot of the navigation mobile apps, and most of the vehicles are equipped with gps units. I suggest the authors borrow ideas from mobile navigation app and remove the dependence of base infrastructures.
Since clouds can provide high-performance computing platforms, the navigation management can be implemented as a powerful cloud-based service. Such an implementation will be able to perform the navigation planning always in a feasible time by offering necessary computing resources according to the size of the navigation management problem and its constraints. %Moreover, this cloud-based architecture is also efficient for collecting or storing runtime traffic informations.

%%%%%%%%%%%%
\subsection{Extendability}
The proposed formal framework for the navigation management synthesis is generic enough to consider further constraints if required. As long as the road network is represented using a set of links connecting the different locations, the model can consider all the destination constraints (\eg delays at the via  points or intermediate point of interests, avoiding the toll roads, highway preference, etc.) toward the destination.

%\vspace{-3pt}
%%%%%%%%%%%%%%%%%%%%%%%%%%%%%%%%%%%%%%%%%%%%%%%%%%%
\section{Related Work}%\vspace{-3pt}
\label{Sec:Related}

In the last few years, many researchers addressed the efficient management of plug-in electric vehicles considering the vehicle-to-grid (V2G) aspects, especially the V2G services like charging, discharging, and frequency regulation services. For example, the works in~\cite{Hutson08} and~\cite{Han10} proposed mechanisms for maximizing the profits from vehicle-to-grid (V2G) services by selling stored electricity to the grid or by participating in frequency regulation. Some other works, such as~\cite{Shi11},~\cite{Mets11}, and our work in~\cite{Rahman13}, developed control algorithms and models for the optimal V2G management. However, none of these works addressed the problem of driving EVs on long trips and scheduling the charging of batteries. %the practical and efficient use of PHEVs as a vehicle especially on highways.

The authors in~\cite{Qin11} addressed the issue of minimizing the EV recharging waiting time through intelligently scheduling recharging activities. The authors theoretically formulated the minimum waiting time for the problem of recharging scheduling. Based on the analysis, they proposed a distributed scheme to optimize the recharging schedule. However, their model is limited only to EVs with a mere objective of minimizing the waiting time in the charging stations only. The model only considers a single highway road and cannot find the navigation/routing path for a complex road network.
Moreover, the work did not consider the alternative use of fuels in the case of hybrid EVs. In this paper, we address the navigation management problem for hybrid electric vehicles for long trips, considering a broader and real-life aspect. %and real-life aspect.

Zhang and Vahidi~\cite{Zhang12} proposed a dynamic programming (DP)-based technique to find strategies to operate the powertrain-based energy management system (EMS) efficiently with the optimal use of (stored) electric charge and gasoline until the next charging station. Bader et al.~\cite{Bader13} developed a DP-based mechanism for predictive real-time energy management with the help of precalculated lookup tables for different points of the powertrain. 
Larsson et al. proposed an energy-optimal route selection mechanism based on historical driving data (logged GPS data) about commuter routes and the EMS optimization~\cite{Larsson14}. The mechanism also precomputes the optimal solutions to the energy management control problems using DP and provide them to a user as a lookup table.
Tribioli et al. proposed a heuristic methodology, utilizing Pontryagin's minimum principle-based optimal control theory, to solve the real-time energy management problem~\cite{Tribioli14}. 
Later, this real-time energy management problem is solved using the estimation distribution algorithm by Qi et al.~\cite{Qi14} and 
the particle swarm optimization algorithm by Chen et al~\cite{Chen15}. Li and Chen~\cite{Liu16} presented a solution for the optimal, real-time energy management of a parallel PHEV (here, the electric charge can be used to drive one axle while the gasoline to drive the other one).
Yonggang et al.~\cite{Liu17} developed an EMS control technique utilizing the GPS/GIS data and the knowledge of road events (trip distance, road grade, altitude, velocity, etc.) to predict the road slope grade and the corresponding electric energy consumption, and thus control the battery usage so that energy is not discharged during uphill.
It is worth mentioning that this strategic operation of the EMS or powertrain for optimal energy management of PHEVs is out of the scope of this work. 
%\comment{Write about the following papers: , and \cite{Yonggang16}.}

%It is worth mentioning the work by 
Raghu K. Ganti et al. developed a navigation service, named GreenGPS~\cite{Raghu11}, for traditional vehicles. GreenGPS provides the most fuel-efficient route between two points, unlike the shortest and fastest routes provided by the traditional navigation tools like Google Maps~\cite{GoogleMap} and MapQuest Maps \cite{MapQuest}. GreenGPS collects the necessary data, which is continuously updated by the participating vehicles and answers queries on the most fuel-efficient route. Later, further results are presented in~\cite{Saremi16} according to an advanced real-world phone-based implementation and deployment of GreenGPS, which considers reliable end-to-end data collection and route recommendation. 
However, GreenGPS cannot work for PHEVs as it does not consider the recharging of these vehicles. It considers mainly the traffic congestion and the stop lights to find out the most fuel-efficient path, which is suitable in urban areas. Moreover, for traveling on highways, GreenGPS will perform almost similarly to a traditional GPS, as there is no stop light and often there is no or minor congestion.   
Nejad et al.~\cite{Nejad17} proposed routing algorithms for PHEVs that account vehicle operating modes (either electric, gasoline, or hybrid) and recommend the most fuel-efficient path by choosing the optimal mode of operation for each road segment during route planning. However, the solution requires to choose a single mode of operation for each segment and to segment the roads appropriately based on the relatively uniform energy efficiency conditions. Moreover, the work does not consider recharging/refueling of these vehicles and thus inapplicable for long trips. 

Our proposed model for long trips is comprehensive (considering PHEV properties), flexible (based on user requirements), and practical (allowing recharging and refueling). The formal modeling of a navigation problem presented in this paper is also unique. Our model is extendable for broader aspects with novel and further requirements.

%%%%%%%%%%%%%%%%%%%%%%%%%%%%%%%%%%%%%%%%%%%%%%%%%%%
\section{Conclusion}
\label{Sec:Conclusion}

Plug-in hybrid vehicles require switching to gasoline or recharging their batteries for long trips. Recharging batteries takes a longer time compared to the refueling of gasoline. Due to these characteristics, a flexible navigation management scheme is required.
In this paper, we have presented an SMT-based formal modeling of the PHEV navigation management problem. Our model offers an optimal management plan that includes the route, along with the potential charging points. This satisfies all the constraints on the fuel cost, the traveling time, and the intermediate points of interest. We have also presented a price-based navigation control technique to achieve better load balancing
for the system. 
%Our model is also flexible and extendable to include additional requirements (\eg time/cost constraints on reaching the intermediate points of interest).
We implemented both of our models, ran simulation experiments using the Z3 SMT solver and evaluated their scalability. 
We observed that the running time of our navigation management model usually lies within few seconds for a road network of a thousand location points, %and similar number of charging stations,
while that of our control model is around 25 seconds for a problem of 20\% charging stations on a road network with 1000 points and 1000 PHEVs.

%%%%%%%%%%%%%%%%%%%%%%%%%%%%%%%%%%%%%%%%%%%%%%%%%%%
% references section
\bibliographystyle{unsrt}
\bibliography{PHEV_Navigation_Rahman_arXiv}

\vspace{36pt} 
  %\vfill
  %\parpic{\includegraphics[width=1.1in,height=1.3in,clip]{Figures/Rahman.pdf}}
  \begin{wrapfigure}{l}{30mm} 
  \vspace{-11pt}
  \includegraphics[width=1.1in,height=1.2in,clip]{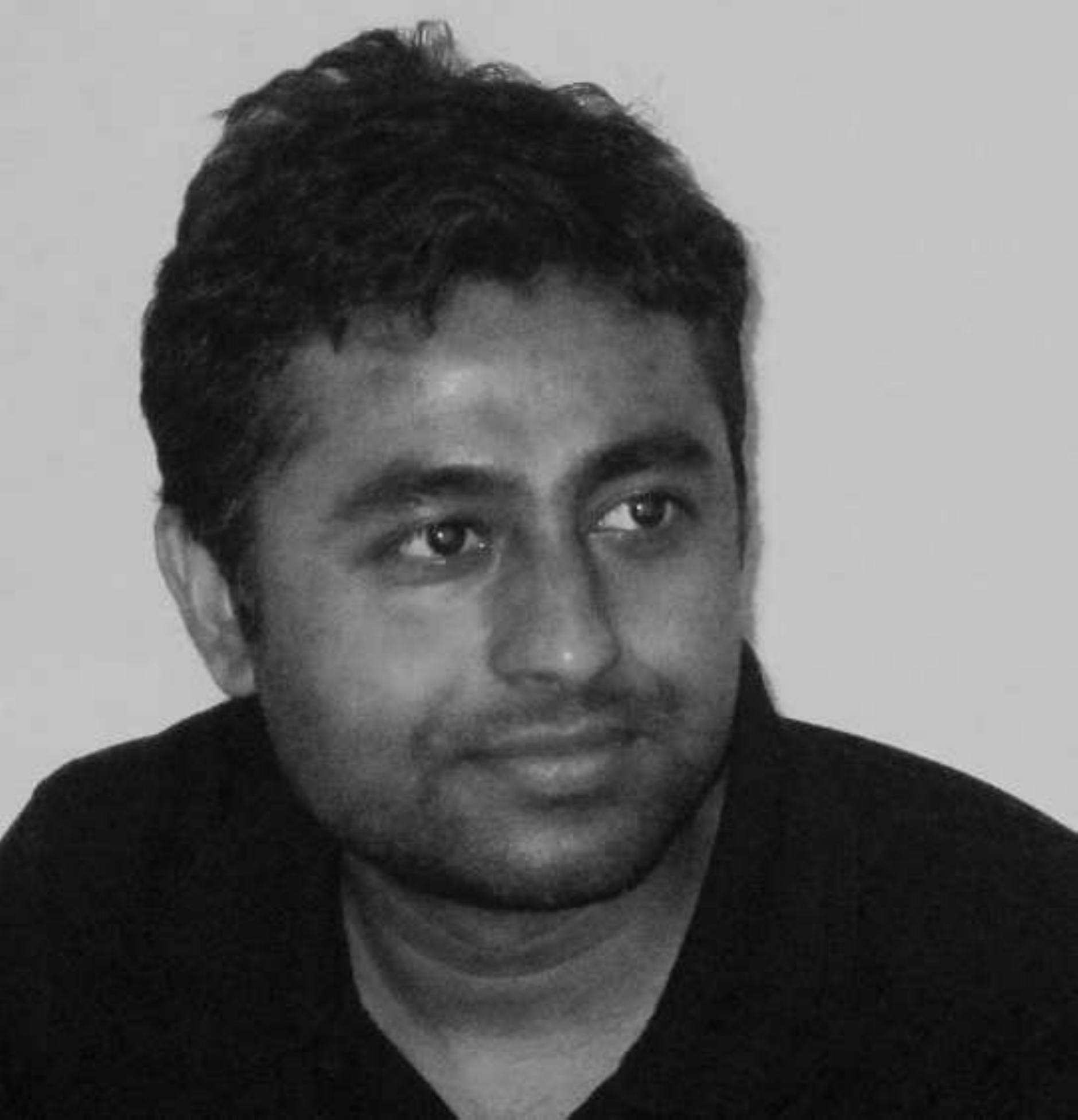}
  \vspace{-6pt}
  \end{wrapfigure}\par
 \noindent{\bf Mohammad Ashiqur Rahman} is an Assistant Professor in the Department of Electrical and Computer Engineering (ECE) at Florida International University (FIU), USA, and leading the Analytics for Cyber Defense (ACyD) Lab at FIU. Before joining FIU, he was an Assistant Professor at Tennessee Tech University. He received the BS and MS degrees in Computer Science and Engineering from Bangladesh University of Engineering and Technology (BUET), Dhaka, in 2004 and 2007, respectively, and obtained the PhD degree in computing and information systems from the University of North Carolina at Charlotte (UNC Charlotte) in 2015. 
 %He is a member of the Center for Energy Systems Research (CESR) and running the Cyber Security, Resiliency, and Sustainability (CSRS) Lab at Tennessee Tech. 
 %
%Rahman's primary research interest covers a wide area of computer networks and communications, within both cyber and cyber-physical systems. 
Rahman's research focus primarily includes computer and information security, risk analysis and security hardening, secure and dependable resource allocation and optimal management, and distributed and parallel computing. 
He has already published over 50 peer-reviewed journals and conference papers. He has also served as a member in the technical programs and organization committees for various IEEE and ACM conferences.

%\vfill
\vspace{12pt}
 \begin{wrapfigure}{l}{30mm} 
   \vspace{-11pt}
  \includegraphics[width=1.1in,height=1.2in,clip]{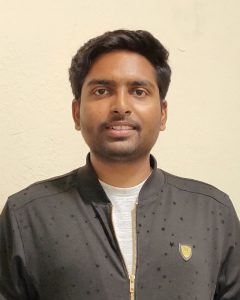}
   \vspace{-6pt}
   \end{wrapfigure}\par
  %\parpic{\includegraphics[width=1.1in,height=1.3in,clip]{Figures/Mehedi.jpg}}
  \noindent{\bf Md Hasan Shahriar} is a PhD student in the Department of ECE at FIU. Earlier, he received his BS in Electrical and Electronics Engineering from BUET, Dhaka, in 2016. After the graduation, he joined Uttara University, Dhaka as a Lecturer in the Department CSE. He also worked as an Assistant Engineer at Electricity Generation Company of Bangladesh (EGCB). Shahriar is interested in the formal analysis of security and resiliency for cyber-physical systems (CPS)/ internet of things (IoT) His current focus areas include smart grid/microgrid and industrial IoT. He is also a member of the ACyD Lab at FIU.

%\vfill

\vspace{12pt}
%\begin{IEEEbiography}[{\includegraphics[width=1in,height=1.25in,clip]{Figures/Ehab.pdf}}] 
\begin{wrapfigure}{l}{30mm} 
  \vspace{-11pt}
  \includegraphics[width=1.1in,height=1.2in,clip]{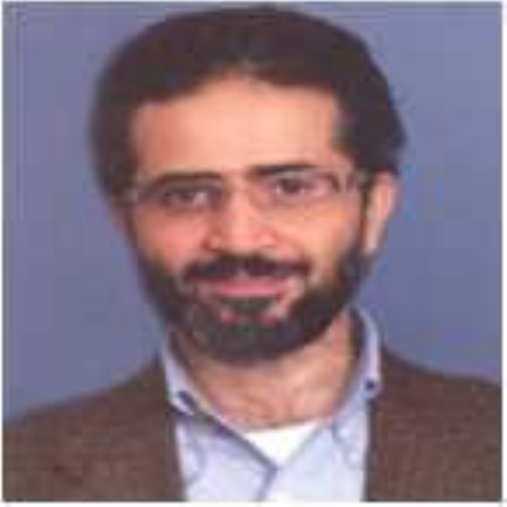}
  \vspace{-6pt}
  \end{wrapfigure}\par
 \noindent{\bf Ehab Al-Shaer} is a Professor and the Director of the Cyber Defense and Network Assurability (CyberDNA) Center in the College of Computing and Informatics at UNC Charlotte. He received his MSc and Ph.D. in Computer Science from the Northeastern University (Boston, MA) and Old Dominion University (Norfolk, VA) in 1998 and 1994 respectively. His primary research areas are network security, security management, fault diagnosis, and network assurability. Prof. Al-Shaer edited/co-edited more than 10 books and book chapters, and published about 200 refereed journals and conferences papers in his area. 
%Prof. Al-Shaer is the General Chair of ACM CCS 2009-2010 and NSF Workshop in Assurable and Usable Security Configuration, August 2008. 
Prof. Al-Shaer also served as a Conference/Workshop Chair and Program Co-chair for a number of well-established conferences/workshops in his area including IM~2007, POLICY~2008, ANM-INFOCOM~2008, ACM~CCS~2009-2010. %He also served as a member in the technical programs and organization committees for many IEEE and ACM conferences. 
%\end{IEEEbiography}

%\vfill

\vspace{12pt}
%\begin{IEEEbiography}[{\includegraphics[width=1in,height=1.25in,clip]{Figures/Quanyan.pdf}}]
\begin{wrapfigure}{l}{30mm} 
  \vspace{-11pt}
  \includegraphics[width=1.1in,height=1.2in,clip]{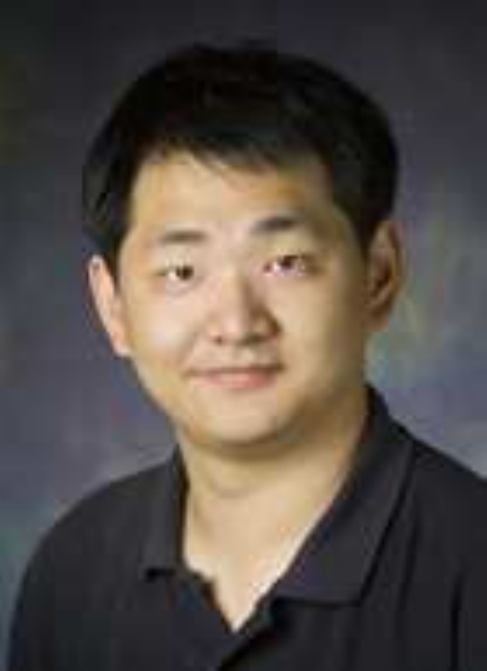}
  \vspace{-6pt}
  \end{wrapfigure}\par
 \noindent{\bf Quanyan Zhu} is an Assistant Professor in the Department of Electrical and Computer Engineering at New York University. He received the B. Eng. in Electrical Engineering from McGill University in 2006, the M.Sc. from University of Toronto in 2008, and the Ph.D. from the University of Illinois at Urbana-Champaign (UIUC) in 2013. %From 2013- 2014, he was a postdoctoral research associate at the Department of Electrical Engineering, Princeton University. 
He is a recipient of many awards including NSERC Canada Graduate Scholarship (CGS), Mavis Future Faculty Fellowships, and NSERC Postdoctoral Fellowship (PDF). He spearheaded the INFOCOM workshop on Communications and Control on Smart Energy Systems (CCSES), the Midwest Workshop on Control and Game Theory (WCGT), and NYU Workshop on Control and Optimization of Network Systems (CONES).
%\end{IEEEbiography}

\vfill

\end{document}